\documentclass{jfm}

\usepackage[T1]{fontenc}
\usepackage[english]{babel}
\usepackage[utf8x]{inputenc}

\usepackage{graphicx}
\usepackage{subcaption}
\usepackage{overpic}

\usepackage{lipsum}
\usepackage{titletoc}
\usepackage{placeins}
\usepackage{amstext}
\usepackage{amsmath,amssymb}

\usepackage{array}
\usepackage{color}
\usepackage{colortbl}

\usepackage[titletoc]{appendix}
\usepackage{esint}

\usepackage{bm}
\usepackage{array,multirow}

\usepackage{hyperref}
\hypersetup{
    colorlinks=true,
    linkcolor=red,
    citecolor=blue,
    filecolor=cyan,
}

\usepackage{xcolor}

\newcommand\avg[1]{\left\langle #1 \right\rangle}

\newcommand\of[1]{\left( #1 \right)}

\newcommand{\de}{\text{\textrm{d}}}
\newcommand{\Tr}{\text{\textrm{Tr}}}
\newcommand{\aH}{\bm{\mathcal{H}}}

\graphicspath{{./figures/}}

\shorttitle{Gauge symmetry for the pressure Hessian}
\shortauthor{M.\ Carbone, M.\ Iovieno, and A.\ D.\ Bragg}

\title{Gauge symmetry and dimensionality reduction of the anisotropic pressure Hessian}

\author{M.\ Carbone\aff{1,2}, M.\ Iovieno\aff{1} \and A.\ D.\ Bragg\aff{2}\corresp{\email{andrew.bragg@duke.edu}}}

\affiliation{\aff{1}Dipartimento di Ingegneria Meccanica e Aerospaziale,
Politecnico di Torino, Corso Duca degli Abruzzi 24, 10129 Torino, Italy
\aff{2}Department of Civil and Environmental Engineering, Duke University, Durham, NC 27708,
USA}

\begin{document}

\maketitle

\begin{abstract}
Analyzing the fluid velocity gradients in a Lagrangian reference frame provides an insightful way to study the small-scale dynamics of turbulent flows, and further insight is provided by considering the equations in the eigenframe of the strain-rate tensor.
The dynamics of the velocity gradient tensor is governed in part by the anisotropic pressure Hessian, which is a non-local functional of the velocity gradient field. This anisotropic pressure Hessian plays a key role in the velocity gradient dynamics, for example in preventing finite-time singularities, but it is difficult to understand and model due to its non-locality and complexity.
In this work a gauge symmetry for the pressure Hessian is introduced to the eigenframe equations of the velocity gradient, such that when the gauge is added to the original pressure Hessian, the dynamics of the eigenframe variables remain unchanged.
We then exploit this gauge symmetry to perform a rank reduction on the three-dimensional anisotropic pressure Hessian, which, remarkably, is possible everywhere in the flow.
The dynamical activity of the newly introduced rank-reduced anisotropic pressure Hessian is confined to two dimensional manifolds in the three dimensional flow, and exhibits striking alignment properties with respect to the strain-rate eigenframe and the vorticity vector.
The dimensionality reduction, together with the strong preferential alignment properties, leads to new dynamical insights for understanding and modelling the role of the anisotropic pressure Hessian in three-dimensional flows.
\end{abstract}


\section{Introduction}

The small-sale dynamics of turbulent flows is governed by highly non-linear and non-local dynamical processes, whose statistics are strongly intermittent in space and time \citep{Yeung2012,Buaria2019}.
Moreover, the strong and intermittent small-scale dynamics can generate coherent structures at larger scales \citep{Majda2001}. Such small-scale dynamics is effectively characterized by the velocity gradient field, rather than the velocity field itself \citep{Tsinober2001}. Consequently, understanding and modelling the velocity gradient dynamics is of singular importance in the study of turbulence and has been the subject of many works in the literature. In particular, the Lagrangian description of the velocity gradient dynamics has proven to be especially fruitful for understanding and modeling \citep{Meneveau2011}.

The equation governing the velocity gradient tensor dynamics along a fluid particle trajectory is easily derived from the Navier-Stokes equation (NSE) but, the equation is unclosed because of the anisotropic/non-local pressure Hessian and viscous terms. Developing closure models for these complex terms requires insight, and this work concentrates on the properties of the anisotropic pressure Hessian.

The pressure field can be expressed as a linear, non-local, functional of the second invariant of the velocity gradient tensor. Therefore, a strategy to infer the statistical properties of the pressure field consists in analyzing how the velocity gradient organizes in space.
A quantitative investigation of the correlation length of the velocity gradient magnitude shows that, in rotation-dominated regions, the pressure field is governed by a dissipation-scale neighbourhood while, in strain-dominated regions, the pressure is determined by an inertial-scale neighbourhood \citep{Vlaykov2019}. However, many works in the literature have shown that the pressure statistics can be described reasonably well by quasi-local approximations \citep{Chevillard2008,Lawson2015}. Indeed, the long-range effects to the pressure field are much smaller than expected due to partial cancellation of the competing contributions of the strain-rate and vorticity magnitude to the second invariant of the velocity gradient \citep{Vlaykov2019}.

The information about the statistics of the pressure field can then be employed to develop closure models for the Lagrangian dynamics of the velocity gradient in turbulence. In the inviscid case, an early closure model by \cite{Vieillefosse1982} has been derived neglecting the non-local/anisotropic part of the pressure Hessian, while retaining its local/isotropic part. This model is usually referred to as the Restricted Euler (RE) model. This model led to important insights, showing the tendency for the intermediate eigenvalue of the strain-rate to be positive, and also the preferred alignment of the vorticity with the intermediate strain-rate eigenvector  \citep{Cantwell1992} as observed in Direct Numerical Simulation (DNS) of isotropic turbulence and homogeneous shear flows \citep{Ashurst1987}.
However, the RE flow exhibits a finite-time singularity for almost all initial conditions, indicating that a realistic model for the velocity gradient should take into account the anisotropic pressure Hessian, in addition to viscous contributions. Indeed, the anisotropic pressure Hessian is considered to play a major role in preventing such finite-time singularities, even for ideal fluids, and it has been analyzed in detail in several works \citep{Ohkitani1993,Nomura1998,Chevillard2008,Vlaykov2019}. 

In an early work, the anisotropic pressure Hessian has been modelled as a stochastic process, independent of the gradient dynamics, and the stochastic differential equations for the velocity gradient have been constructed to satisfy isotropy constraints and empirical constraints as the log-normality of the dissipation rate \citep{Girimaji1990a}. A more advanced phenomenological and stochastic model was constructed in \cite{Chertkov1991} by analyzing the Lagrangian dynamics using four tracer trajectories, forming a tetrad.  The tetrad can be used to construct a scale-dependent filtered velocity gradient  \citep{Naso2005}  and the closure of the model involves a  direct relation between the local pressure and the velocity gradient on the tetrad. 
The tetrad model provided a phenomenological basis for understanding how the anisotropic pressure Hessian acts to reduce non-linearity in the flow, a property that also emerges in more systematic closures for the pressure Hessian based on Gaussian random fields \citep{Wilczek2014}.

The deformation history of a fluid particle in the flow has been employed to model the anisotropic pressure Hessian and viscous terms using Lagrangian coordinate closures \citep{Chevillard2006}. In this model, only information on the recent fluid deformation (RFD) is retained, that is, the dynamics is affected by times up to the Kolmogorov timescale, $\tau_\eta$, in the past. A phenomenological closure is then constructed assuming that at a time $\tau_\eta$ in the past, the Lagrangian pressure Hessian was isotropic. This model does not exhibit the singularity associated with the RE, and was shown to capture many of the non-trivial features of the velocity gradient dynamics that are observed in experiments and Direct Numerical Simulations of the NSE. However, it displays unphysical behaviour for flows at large Reynolds number. A critical comparison with DNS data \citep{Chevillard2008} showed that while the closure model presented in \cite{Chevillard2006} can reproduce some of the non-trivial velocity gradient dynamics, it misses some important features of the pressure Hessian dynamics and statistical geometry in the flow.

\cite{Wilczek2014} proposed a closure for the Lagrangian velocity gradient equation by assuming that the velocity is a random field with Gaussian statistics. Closed expressions for the pressure Hessian and viscous terms conditioned on the velocity gradient are obtained by means of the characteristic functional of the Gaussian velocity field. The model produces qualitatively good results but, owing to the Gaussian assumption, it leads to quantitative predictions that are not in full agreement with DNS data. Therefore, to correct this aspect, the authors modified the closure such that the mathematical structure was retained, but the coefficients appearing in the model were prescribed using DNS data. This led to significant improvements, since the model provides interesting insights into the role of the anisotopic pressure Hessian in preventing the singularities arising in the RE. However, 
the enhanced model did not satisfy the kinematic relations for incompressible and isotropic flows \citep{Betchov1956}.

Another model has been developed by \cite{Johnson2016}, who combined the closure modeling ideas by both \cite{Chevillard2006} and \cite{Wilczek2014}. This model leads to improvements compared with the two models on which it is based, and it is formulated in such a way that by construction the model satisfies the kinematic relations of \cite{Betchov1956}. However, a quantitative comparison with DNS data revealed some shortcomings in the ability of the model to properly capture the intermittency of the flow. Moreover, it runs into difficulties for high Reynolds number flows, like that of \cite{Chevillard2006} from which it has been partly derived. The capability to reproduce intermittency and high-Reynolds number flow features is a major challenge for velocity gradient models. A recent development of velocity gradient models, based on a multiscale refined self-similarity hypothesis, proposed by \cite{Johnson2017}, seems to remove the Reynolds number limitations (at least in the sense that the model does not break down at high Reynolds numbers).

In summary, while significant progress has been made since the initial modelling efforts of \cite{Vieillefosse1982,Vieillefosse1984}, much remains to be done. A major difficulty in developing accurate closure approximations for the Lagrangian velocity gradient equation is that the dynamical effects  of the anisotropic/non-local pressure Hessian on the flow are not yet fully understood and are difficult to approximate using simple closure ideas.
This fact is the motivation behind the present work which aims to improve the understanding of the anisotropic pressure Hessian, and in particular, its statistical geometry relative to the strain-rate and vorticity fields.
In the following, we present what appears to be a previously unrecognized gauge symmetry for the pressure Hessian, such that when this gauge is added to the pressure Hessian, the invariant dynamics of the velocity gradient tensor remains unchanged. We then exploit this gauge symmetry to perform a rank reduction on the anisotropic pressure Hessian.
Remarkably, this rank reduction can be performed everywhere in the turbulent flow, and produces the newly introduced rank-reduced anisotropic pressure Hessian which lives on a two-dimensional manifold and exhibits striking alignment properties with respect to the strain-rate eigenframe and the vorticity vector. This dimensionality reduction, together with evident preferential alignments of the rank-reduced anisotropic pressure Hessian has implications in the understanding and modelling of turbulent flows.
\section{Theory}

In this Section the gauge symmetry for the invariants dynamics is derived from the equations for the velocity gradient written in the strain-rate eigenframe. The gauge is then  exploited to reduce the rank of the anisotropic pressure Hessian obtaining a rank-reduced anisotropic pressure Hessian which is a two-dimensional object embedded in a three-dimensional space.

\subsection{Equations for the fluid velocity gradient in the strain-rate eigenframe}

The three-dimensional flow of a Newtonian and incompressible fluid with unitary density is described by the  Navier-Stokes equations
\begin{equation}
D_t\bm{u} \equiv \partial_t\bm{u}  + (\bm{u\cdot}\nabla) \bm{u} = -\nabla P + \nu\nabla^2\bm{u},\quad \nabla\bm{\cdot}\bm{u} = 0,
\label{eq_NS}
\end{equation}
where $\bm{u}(t,\bm{x})$, $P(t,\bm{x})$ are the fluid velocity and pressure fields and $\nu$ is the kinematic viscosity. By taking the gradient of \eqref{eq_NS}, the following equation for the velocity gradient tensor is obtained
\begin{align}
D_t\bm{A}  = -\bm{A\cdot A} - \bm{H} + \nu \nabla^2 \bm{A},\quad \Tr(\bm{A} )&= 0,
\label{eq_grad}
\end{align}
where $\bm{A}\equiv \bm{\nabla u}$ is the velocity gradient, and $\bm{H}\equiv \bm{\nabla\nabla}P$ is the pressure Hessian.
The pressure and viscous terms in equation \eqref{eq_grad} are not in closed form, since they cannot be expressed in terms of the velocity gradient along the fluid particle trajectory, $\bm{A}(t,\bm{x}(t))$. Models are necessary to define those terms and reliable modelling of them requires an understanding of their dynamical and statistical properties \citep{Meneveau2011}.

The tensor $\bm{A}$ is  decomposed into its symmetric and anti-symmetric part, namely the strain-rate $\bm{S} \equiv (\bm{A} + \bm{A}^\top)/2$, and the rate-of-rotation $\bm{R} \equiv (\bm{A} - \bm{A}^\top)/2$, whose components are related to the vorticity $\bm{\omega}\equiv\bm{\nabla}\times\bm{u}$ as $R_{ij}=\epsilon_{ikj}\omega_k/2$.
Using equation \eqref{eq_grad} the equations for $\bm{S}$ and $\bm{\omega}$ are obtained, and it is insightful to write these in the eigenframe of $\bm{S}$. The eigenvectors $\bm{v}_i$ of the strain-rate satisfy $\bm{v}_i\bm{\cdot v}_j=\delta_{ij}$, where $\delta_{ij}$ is the Kronecker delta, and thus define an orthonormal basis. The strain-rate eigenvectors remain orthogonal so that the strain-rate basis undergoes rigid body rotation only, with rotation rate $\bm{w}$,
\begin{equation}
D_t\bm{v}_i = \bm{w}\times\bm{v}_i.
\end{equation}
The equations for the velocity gradient in the strain-rate eigenframe read
\begin{align}
\sum_{j=1}^3 \lambda_j &= 0 \label{eq_cont_eigen}\\
D_t{\lambda_{i}} &=  -\lambda^2_{i} + \frac{1}{4}\left( \omega^2 - \widetilde{\omega}_i^2\right) - \widetilde{H}_{i(i)} +\widetilde{ \nu\nabla^2 S_{i(i)}},\label{eq_lambda}\\
\widetilde{W}_{ij}\left(\lambda_{(j)}-\lambda_{(i)}\right) &= -\frac{1}{4} \widetilde{\omega}_i\widetilde{\omega}_j - \widetilde{H}_{ij} + \widetilde{\nu\nabla^2{S}_{ij}}, \; j \ne i,\label{eq_alg}\\
D_t{\widetilde{\omega}_i} &= \lambda_{(i)}\widetilde{\omega}_i - \widetilde{W}_{ij}\widetilde{\omega}_j +\widetilde{ \nu\nabla^2\omega_i},\textrm{ for $i=1,2,3$}\label{eq_omega}
\end{align}
where $\lambda_i$ are the strain-rate eigenvalues, the tilde indicates tensors components in the strain-rate eigenframe, so that $\widetilde{\omega}_i=\bm{v}_i\bm{\cdot\omega}$ and $\widetilde{H}_{ij}=\bm{v}_i\bm{\cdot H \cdot v}_j$ and ${\omega}^2\equiv\widetilde{\omega}_i\widetilde{\omega}_i$.
In these equations, the indexes in brackets are not contracted.
The anti-symmetric tensor $\bm{W}$ is related to the eigenframe angular velocity $\bm{w}$ through
\begin{equation}
W_{ij}=\epsilon_{ikj}w_k
\end{equation}
and $\widetilde{W}_{ij}$ are the components of $\bm{W}$ in the strain-rate eigenframe. The eigenframe equations \eqref{eq_cont_eigen}, \eqref{eq_lambda}, \eqref{eq_alg}, \eqref{eq_omega} allow to sort out the interaction between local strain and vorticity and have been studied in detail \citep{Vieillefosse1982,Dresselhaus1992,Nomura1998}.
\subsection{A new symmetry for the dynamics of the velocity gradient invariants}

The eigenframe equations satisfy basic symmetries.
They are naturally invariant under the transformation $\widetilde{\omega}_i \to-\widetilde{\omega}_i$, since the eigenvectors are only defined up to an arbitrary sign.
The inviscid equations are also invariant under time reversal $t\rightarrow -t$.
However, the equations also possess another kind of symmetry that does not appear to have been previously recognized.
That new symmetry arises from the fact that in the equation governing $\widetilde{\omega}_i$, the strain-rate eigenrame rotation rate $\bm{w}$ only enters through the cross product $\widetilde{W}_{ij}\widetilde{\omega}_j$ and therefore its component along the vorticity direction, $\bm{w\cdot\omega}$, does not affect in any way the time evolution of the velocity gradient invariants.
In order to show this fact we first define the transformation
\begin{eqnarray}
\bm{W} \to \bm{W} + \gamma\bm{R},
\label{eq_gauge_W}
\end{eqnarray}
that corresponds to adding to the rotation-rate of the strain-rate eigenframe an additional rotation about the vorticity axis at rate $\gamma\omega/2$, where $\gamma(t,\bm{x})$ is a non-dimensional scalar field. If we introduce the transformation \eqref{eq_gauge_W} into the eigenframe equations, the equation governing the strain-rate eigenvalues \eqref{eq_lambda} and the vorticity components in the strain-rate eigenframe \eqref{eq_omega} remain unchanged. Indeed the equation for $\lambda_i$ is not affected by the transformation \eqref{eq_gauge_W} since it does not contain $\bm{W}$. The equation for $\widetilde{\omega}_i$ is also unaffected  since by definition $\bm{R\cdot\omega}=\bm{0}$ and, therefore,
\begin{equation}
D_t{\widetilde{\omega}_i} = \lambda_{(i)}\widetilde{\omega}_i - \left[\widetilde{W}_{ij}+\gamma \widetilde{R}_{ij}\right]\widetilde{\omega}_j + \widetilde{ \nu\nabla^2\omega_i}= \lambda_{(i)}\widetilde{\omega}_i - \widetilde{W}_{ij}\widetilde{\omega}_j + \widetilde{ \nu\nabla^2\omega_i}.
\label{eq_omega_mod}
\end{equation}
On the other hand, the off-diagonal algebraic equation \eqref{eq_alg} becomes
\begin{equation}
\widetilde{W}_{ij}\left(\lambda_{(j)}-\lambda_{(i)}\right) = -\frac{1}{4} \widetilde{\omega}_i\widetilde{\omega}_j - \widetilde{H}_{ij} -\gamma \widetilde{R}_{ij}\left(\lambda_{(j)}-\lambda_{(i)}\right) +  \widetilde{\nu\nabla^2{S}_{ij}}, \; j\ne i.
\label{eq_off_diag_mod}
\end{equation}
This equation is not invariant under the transformation \eqref{eq_gauge_W}. However, while this changes the orientation of the strain-rate eigenframe with respect to a fixed, arbitrary, reference frame, it does not affect either $\lambda_i$ or $\widetilde{\omega}_i$.
Therefore, the transformation $\bm{W} \to \bm{W} + \gamma\bm{R}$ corresponds to a symmetry for the invariants of the velocity gradient tensor, that can be expressed in terms of $\lambda_i$ or $\widetilde{\omega}_i$. For example, the second and third invariants of the velocity gradient tensor can be written as
\begin{align}
Q = -\sum_i\lambda_i^2/2+\sum_i\widetilde{\omega}_i^2/4, && 
R = -\sum_i\lambda_i^3/3-\sum_i\lambda_i\widetilde{\omega}_i^2/4.
\label{eq_def_RQ}
\end{align}
It is important to note, however, that multi-time or multi-point invariants of the velocity gradients are not in general invariant under the gauge transformation. For example, $\bm{S}(t,\bm{x}(t))\bm{:S}(t',\bm{x}(t'))$ is affected by the gauge transformation since the transformation arbitrarily modifies the relative orientations of the eigenframes of $\bm{S}(t,\bm{x}(t))$ and $\bm{S}(t',\bm{x}(t'))$. Nevertheless, multi-time or multi-point products of $\lambda_i$ or $\widetilde{\omega}_i$ are invariant under the gauge transformation. In this paper, we focus on single-point and single-time quantities.

\subsection{Gauge symmetry for the anisotropic pressure Hessian}
The anisotropic/non-local pressure Hessian is defined as
\begin{align}
\aH\equiv\bm{H}-\frac{1}{3}\mathbf{I}\Tr(\bm{H})=\bm{H}+\frac{1}{3}\mathbf{I}(\bm{A : A}),
\end{align}
where $\mathbf{I}$ is the three-dimensional identity matrix. This anisotropic pressure Hessian satisfies $\Tr(\aH)=0$ and contains all of the non-local part of $\bm{H}$. It is also important to notice that its non-local dependence on the flow field is only through the second invariant of the velocity gradient $Q$ \citep{Majda2001}.
The invariance of the eigenframe dynamics under the transformation $\bm{W} \to \bm{W} + \gamma\bm{R}$ is interpreted as a gauge symmetry for $\aH$. That is, the term $\gamma \widetilde{R}_{ij}\left(\lambda_{(j)}-\lambda_{(i)}\right)$ in equation \eqref{eq_off_diag_mod} is added to $\widetilde{\mathcal{H}}_{ij}$ defining $\aH_\gamma=\aH+\delta \aH$, without affecting the eigenframe dynamics, which is described through $\lambda_i$ and $\widetilde{\omega}_j$.
In particular, the gauge term
\begin{equation}
\delta\aH = \gamma \sum_{i,j}\widetilde{R}_{ij}\left(\lambda_{j}-\lambda_{i}\right)\bm{v}_i\bm{v}_j^\top
\end{equation}
is the commutator of anti-symmetric and symmetric part of the velocity gradient
\begin{equation}
\delta\aH = \gamma\left[\bm{R},\bm{S}\right],
\label{eq_gauge_comm}
\end{equation}
where $\left[\bm{R},\bm{S}\right]\equiv\bm{R\cdot S}-\bm{S\cdot R}$. Then, the gauge symmetry consists in the fact that the single-point and single-time Lagrangian dynamics of the velocity gradient invariants is identical when $\aH$ is replaced by
\begin{equation}
\aH_\gamma = \aH + \gamma[\bm{R},\bm{S}].
\label{eq_H_gamma}
\end{equation}
The gauge symmetry holds for all real and finite multiplier $\gamma(t,\bm{x})$, which at this stage is still undetermined. 

It is interesting to note that a term identical to that in equation \eqref{eq_gauge_comm} also arises from a closure of the pressure Hessian assuming a random velocity field with Gaussian statistics \citep{Wilczek2014}. In the framework of the Gaussian closure, the coefficient of $[\bm{R},\bm{S}]$ is the only one that requires specific knowledge of the spatial structure of the flow and must be prescribed by phenomenological closure hypothesis, while all other coefficients of the model can be determined exactly. However, our analysis implies that the ability of the Gaussian closure to predict the invariants of the velocity gradient tensor will not be impacted by the phenomenological closure hypothesis, since its contribution in the dynamics corresponds to the gauge term in equation \eqref{eq_gauge_comm} that does not affect the velocity gradient invariants.
\subsection{Using the gauge symmetry for dimensionality reduction}\label{RR}
While any finite and real $\gamma$ provides a suitable $\aH_\gamma$, there may exist certain choices of $\gamma$ that generate representations of $\aH_\gamma$ that live on a lower dimensional manifold in the system (in the sense that some of its eigenvalues are zero). If such configurations exist and are common, this could significantly aid the understanding and modelling of the anisotropic pressure Hessian in the turbulence dynamics. To seek for such lower dimensional configurations is equivalent to seek for configurations in which a rank-reduction on $\aH_\gamma$ can be performed. We denote such rank-reduced forms of $\aH_\gamma$ by $\aH_\gamma^*$. Notice that $\mathrm{rk}(\aH_\gamma^*)=1$ is not possible since $\Tr(\aH_\gamma)=0$, and therefore either $\mathrm{rk}(\aH_\gamma^*)=2$ or $\aH_\gamma^*=\bm{0}$.

In seeking for lower dimensional representations, when $\aH$ is singular the gauge term is not needed as $\aH$ already lives on a lower dimensional manifold and we take $\aH_\gamma^*=\aH$, corresponding to the choice $\gamma=0$.
On the other hand, when $\aH$ is not singular we seek for a non-zero vector $\bm{z}_2$ such that $\aH_\gamma^*\bm{\cdot z}_2=\bm{0}$, where $\bm{z}_2$ corresponds to the eigenvector of $\aH_\gamma^*$ associated with its zero (and intermediate) eigenvalue. This is equivalent to the generalized eigenvalue problem
$\det\left(\aH_\gamma^*\right)=0$, that is,
\begin{align}
\det\left(\mathbf{I} + \gamma\aH^{-1}\left[\bm{R},\bm{S}\right]\right) = 0.
\label{eq_gen_evp}
\end{align}
Notice that $\aH$ can be safely inverted in equation \eqref{eq_gen_evp}, since the case of singular $\aH$ has been already taken into account and corresponds to $\gamma=0$.
If there exist finite and real values for $\gamma$ that solve equation \eqref{eq_gen_evp}, then those values of $\gamma$ generate a rank-two $\aH_\gamma^*$. Defining $\bm{\mathcal{E}}\equiv\aH^{-1}\left[\bm{R},\bm{S}\right]$, the characteristic equation governing $\xi\equiv -1/\gamma$ reads
\begin{align}
\xi^3 -c\xi^2-b\xi - a= 0,
\label{eq_psi}
\end{align}
with coefficients $a,b,c\in \mathbb{R}$ given by
\begin{align}
a\equiv \det(\bm{\mathcal{E}}), && b\equiv \frac{1}{2}\left(\bm{\mathcal{E}:\mathcal{E}}-\Tr(\bm{\mathcal{E}})\Tr(\bm{\mathcal{E}})\right), && c\equiv \Tr(\bm{\mathcal{E}}).
\label{abc}
\end{align}
The properties of the roots of \eqref{eq_psi} are determined by the discriminant of the polynomial
\begin{align}
\mu\equiv b^2 c^2 + 4b^3 - 4c^3a - 27a^2 - 18abc.
\label{disc}
\end{align}
When $\mu=0$, all of the roots of \eqref{eq_psi} are real and at least two are equal, when $\mu>0$ there are three distinct real roots, and when $\mu<0$ there is one real root and two complex conjugate roots. In every case, there is at least one real root since all the coefficients are real and the degree of the characteristic polynomial is odd. Provided that $a\neq 0$, a real and finite  $\gamma\equiv -1/\xi$  exists. When $a=0$, a real and finite $\gamma$ may or may not exist according to the value of the discriminant $\mu$.
This shows that configurations where a rank-two $\aH_\gamma$ does not exist, that is, the pressure Hessian is intrisically three-dimensional, may only occur when $a= 0$. Interestingly, 
$a\equiv\det\aH\det[\bm{R},\bm{S}]$
and, since by hypothesis $\det\aH\ne 0$ the rank reduction of the anisotropic pressure Hessian may not be performed where $\det[\bm{R},\bm{S}]=0$. The determinant of the commutator is
\begin{equation}
\det[\bm{R},\bm{S}] = \frac{1}{4}(\lambda_2-\lambda_1)(\lambda_3-\lambda_2)(\lambda_1-\lambda_3)
\widetilde{\omega}_1\widetilde{\omega}_2\widetilde{\omega}_3,
\label{eq_detC}
\end{equation}
so that, when either one or more of the vorticity components in the strain-rate eigenframe is zero, and/or the straining-rate configuration is axisymmetric, a singular $\aH_\gamma$ may not exist. However, since $\bm{S}$ and $\bm{\omega}$ have continuous probability distributions, then the probability that $\det[\bm{R},\bm{S}] =0$ is in fact zero. Therefore, the rank reduction of $\aH_\gamma$ should be possible everywhere in the flow.

Configurations in which  multiple rank-reduced anisotropic pressure Hessian can be defined at the same point, that is, there exist more than a single real and finite multiplier $\gamma$, admit an additional discrete symmetry which allows different $\aH_\gamma^*$ to generate the same dynamics of the velocity gradient invariants. We fix this additional gauge by choosing $\gamma$ that provides the maximum alignment between the intermediate eigenvector of the rank-reduced anisotropic pressure Hessian and the vorticity. As it will be shown in \S\ref{Results}, this is justified on the basis of the numerical results, which indicate a marked preferential alignment of the intermediate eigenvector of the rank-reduced anisotropic pressure Hessian with the vorticity.

The rank reduction of the anisotropic pressure Hessian, defined through equation \eqref{eq_gen_evp}, allows for a noticeable reduction of the complexity of the anisotropic pressure Hessian leading to a better understanding of its dynamical effects. Indeed, the fully three-dimensional anisotropic pressure Hessian is specified by five real numbers, being a square matrix of size three. In particular, it takes two numbers to specify the normalized eigenvector $\bm{y}_1$, one additional number for $\bm{y}_2$ (then $\bm{y}_3$ is automatically determined) and two more numbers for the independent eigenvalues $\varphi_1$ and  $\varphi_3$ (since $\sum_i\varphi_i=0$). Therefore, the anisotropic pressure Hessian can be written as
\begin{equation}
\aH = \sum_{i=1}^3\varphi_i\bm{y}_i\bm{y}_i^\top.
\end{equation}
We keep the standard convention $\varphi_1\ge\varphi_2\ge\varphi_3$. On the other hand, the rank-reduced anisotropic pressure Hessian is specified by only four real numbers. Indeed it is a traceless and singular square matrix of size three. In particular, it takes two numbers to specify the plane orthogonal to the normalized eigenvector $\bm{z}_2$ an additional number to specify the orientation of $\bm{z}_1$ on the plane orthogonal to $\bm{z}_2$ (then $\bm{z}_3$ is determined) and a number for the single independent eigenvalue $\psi$.
Therefore, the rank-reduced anisotropic pressure Hessian can be written as
\begin{equation}
\aH_\gamma^* = \psi \left(\bm{z}_1\bm{z}_1^\top - \bm{z}_3\bm{z}_3^\top\right)
\end{equation}
since the intemediate eigenvector is identically zero and the others satisfy $\psi_1=-\psi_3=\psi$ and $\psi\ge 0$. The pressure Hessian lives locally on the plane $\Pi_2$ orthogonal to $\bm{z}_2$, which is the tangent space to a more complex manifold. The tensor $\aH_\gamma^*$ acts on a generic vector $\bm{q}$ amplifying its component along $\bm{z}_1$, cancelling its component along $\bm{z}_2$ and amplifying and flipping its component along $\bm{z}_3$. The rank-reduced anisotropic pressure Hessian is effective only on the plane $\Pi_2$.
The eigenvalue of the rank-reduced anisotropic pressure Hessian can be related to the full anisotropic pressure Hessian and the vorticity since
$
{\bm{\omega^\top \cdot \aH\cdot\omega} = \bm{\omega^\top\cdot\aH}_\gamma\bm{\cdot\omega}}
$
which implies
\begin{equation}
\psi = \frac{\sum_i\varphi_i(\bm{\omega\cdot y}_i)^2}{(\bm{\omega\cdot z}_1)^2-(\bm{\omega\cdot z}_3)^2}.
\label{eq_rel_psi1}
\end{equation}
Moreover, the tensors $\aH$ and $\aH_\gamma$ satisfy the relation
$
{\bm{\omega^\top \cdot S\cdot\aH\cdot\omega} = \bm{\omega^\top \cdot S\cdot\aH}_\gamma\bm{\cdot\omega}}
$
which yields another equation for the eigenvalue $\psi$,
\begin{equation}
\psi = \frac
{\sum_i\varphi_i\bm{\omega\cdot y}_i(\bm{S\cdot\omega})\bm{\cdot y}_i}
{\bm{\omega\cdot}\left[ \bm{z}_1 (\bm{S\cdot\omega})\bm{\cdot z}_1 - \bm{z}_3 (\bm{S\cdot\omega})\bm{\cdot z}_3\right]}.
\label{eq_rel_psi2}
\end{equation}
Equation \eqref{eq_rel_psi1} shows that a perfect alignment between $\bm{z}_2$ and $\bm{\omega}$ would result in an infinitely large $\psi$, unless the anisotropic pressure Hessian fulfills the condition $\bm{\omega^\top \cdot\aH \cdot\omega}=0$. For example, such a peculiar configuration occurs when the flow is exactly two-dimensional, for which $\aH_\gamma^*=\aH$. In general, a large eigenvalue $\psi$ corresponds to strong alignment between $\bm{z}_2$ and $\bm{\omega}$, as it will be discussed in \S\ref{Results}.

\begin{figure}
\centering
\begin{overpic}[width=.49\textwidth]{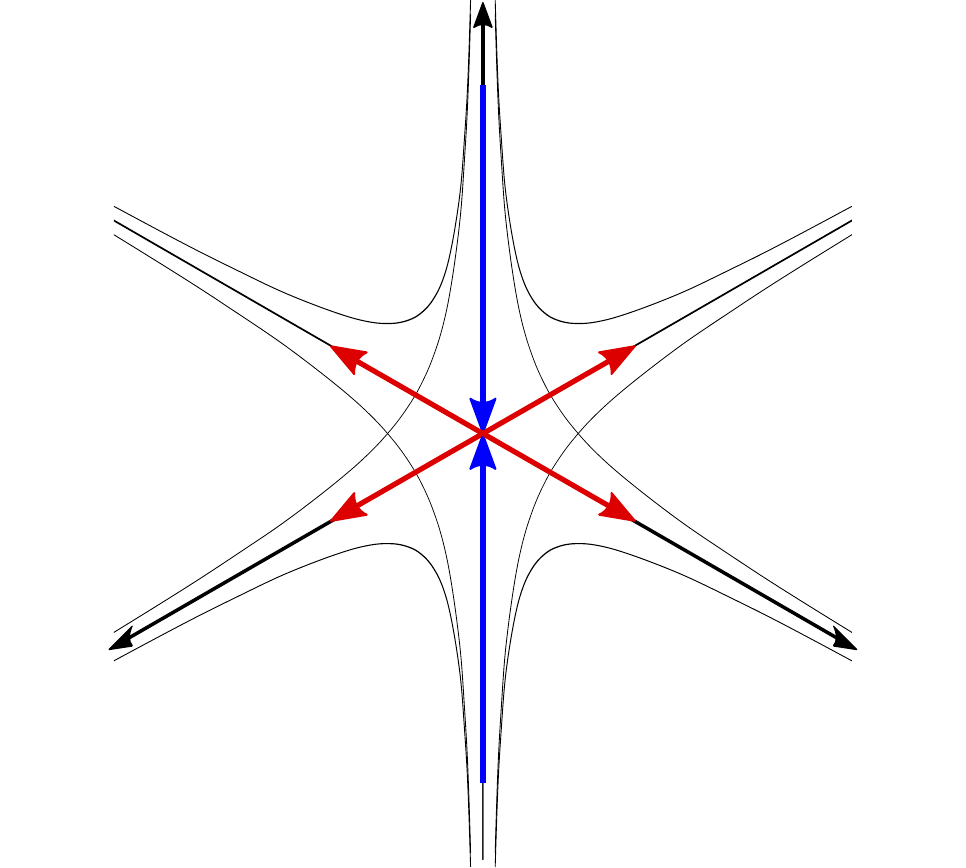}
\put(0,75) {$-\left[\bm{S \cdot S} - \frac{1}{3}(\bm{S : S})\mathbf{I}\right]$}
\put(12,30) {$\bm{v}_1$}
\put(80,30) {$\bm{v}_2$}
\put(54,86) {$\bm{v}_3$}
\put(25,84) {(a)}
\end{overpic}
\begin{overpic}[width=.49\textwidth]{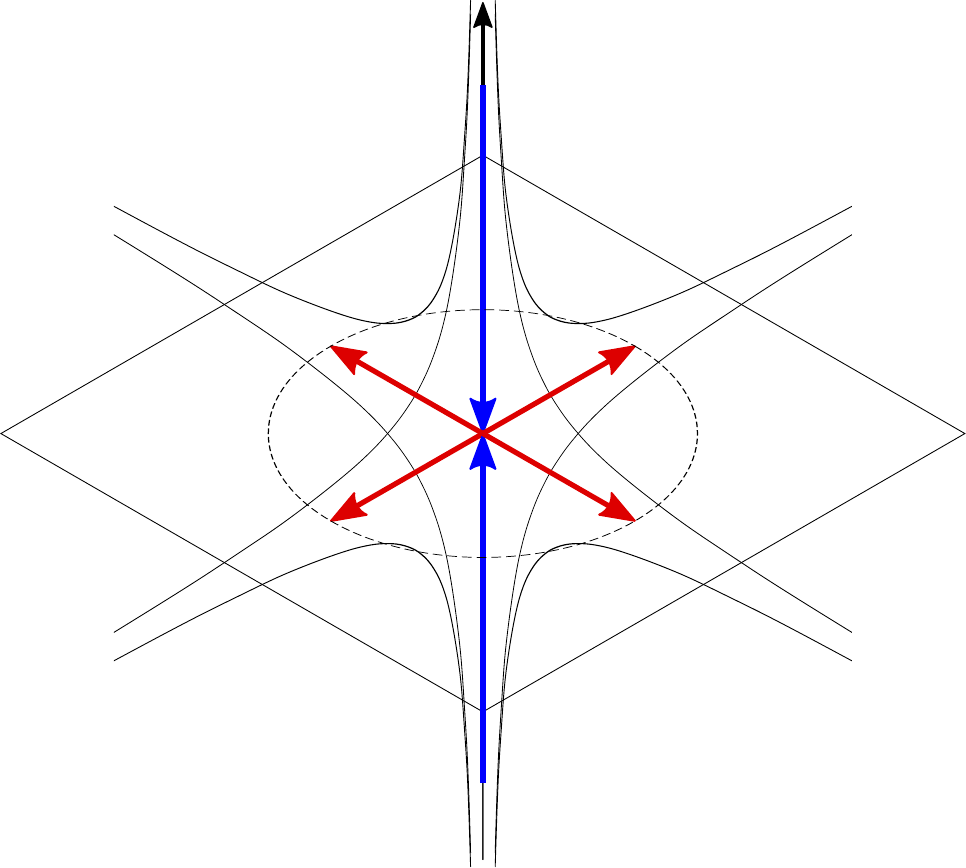}
\put(-2,75) {$-\left[\bm{R \cdot R} - \frac{1}{3}(\bm{R : R})\mathbf{I}\right]$}
\put(80,46) {$\Pi_{\bm{\omega}}$}
\put(54,86) {$\bm{\omega}$}
\put(25,84) {(b)}
\end{overpic}
\vfill
\begin{overpic}[width=.49\textwidth]{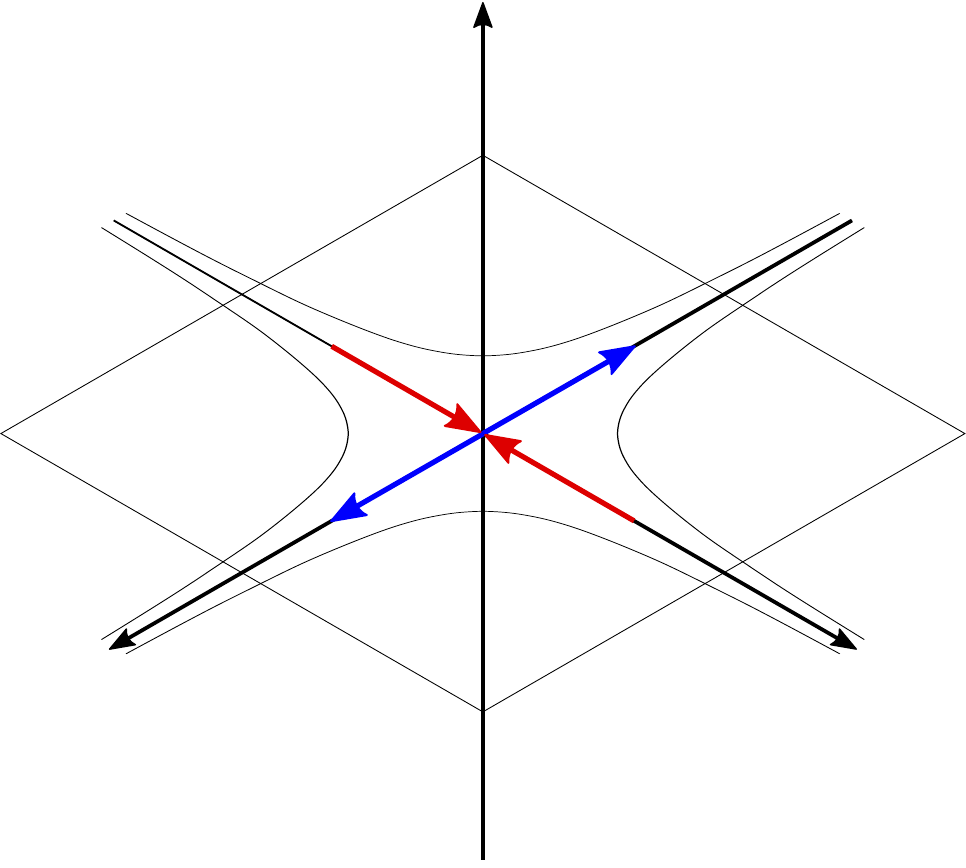}
\put(30,75) {$-\aH_\gamma^*$}
\put(80,46) {$\Pi_2$}
\put(12,30) {$\bm{z}_3$}
\put(54,86) {$\bm{z}_2$}
\put(80,30) {$\bm{z}_1$}
\put(25,84) {(c)}
\end{overpic}
\begin{overpic}[width=.49\textwidth]{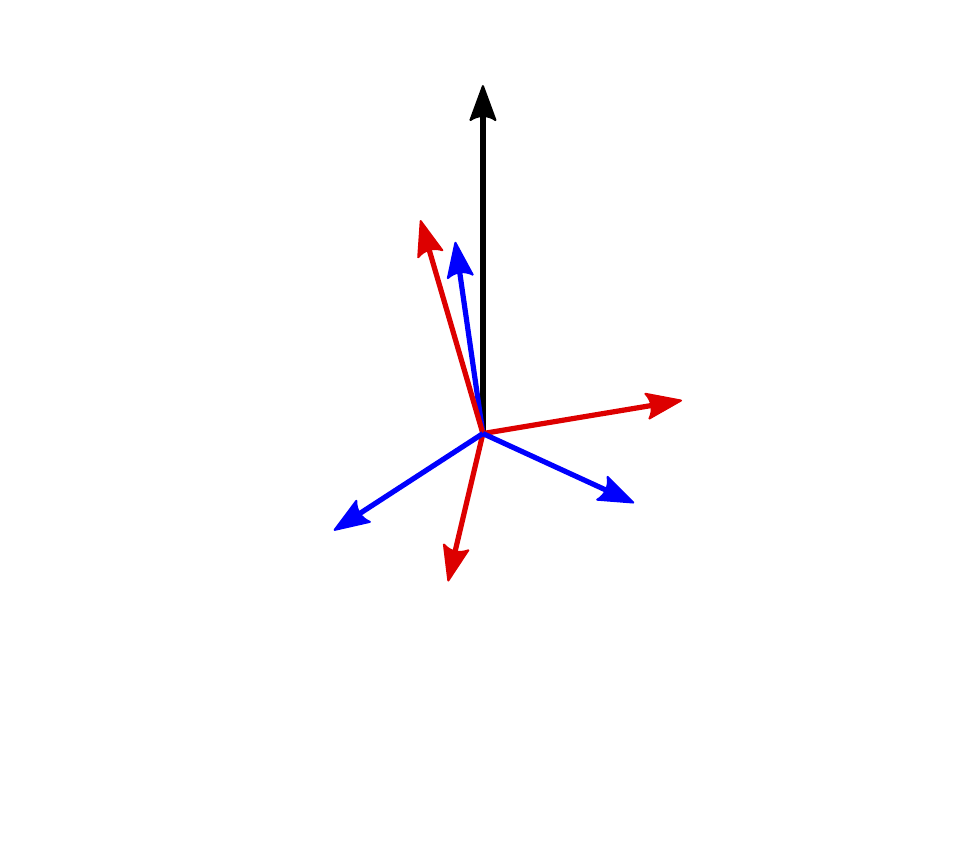}
\put(54,78) {$\bm{\omega}$}
\put(34,40) {$\bm{z}_3$}
\put(64,40) {$\bm{z}_1$}
\put(68,50) {$\bm{v}_1$}
\put(40,30) {$\bm{v}_3$}
\put(25,84) {(d)}
\end{overpic}
\caption{Schematic representation of contribution of the terms on the right hand side of equation \eqref{eq_lambda_iso}. (a) Strain term 
$-\left[\bm{S \cdot S} - \mathbf{I}(\bm{S : S})/3\right]$
for the typical configuration $\lambda_1 =\lambda_2=-\lambda_3/2$. (b) Rotation term
$-\left[\bm{R \cdot R} - \mathbf{I}(\bm{R : R})/3\right]$
which isotropically produces stretching rate along the plane orthogonal to $\bm{\omega}$ and a compression parallel to $\bm{\omega}$. (c) Rank-reduced anisotropic pressure Hessian
$-\aH_\gamma^*$
which produces straining along the $\bm{z}_3$ direction, and hinders it along the $\bm{z}_1$ direction. (d) Typical configuration for the relative orientation of strain-rate eigenframe, vorticity and rank-reduced anisotropic pressure Hessian eigenframe. }
\label{fig_scheme}
\end{figure}
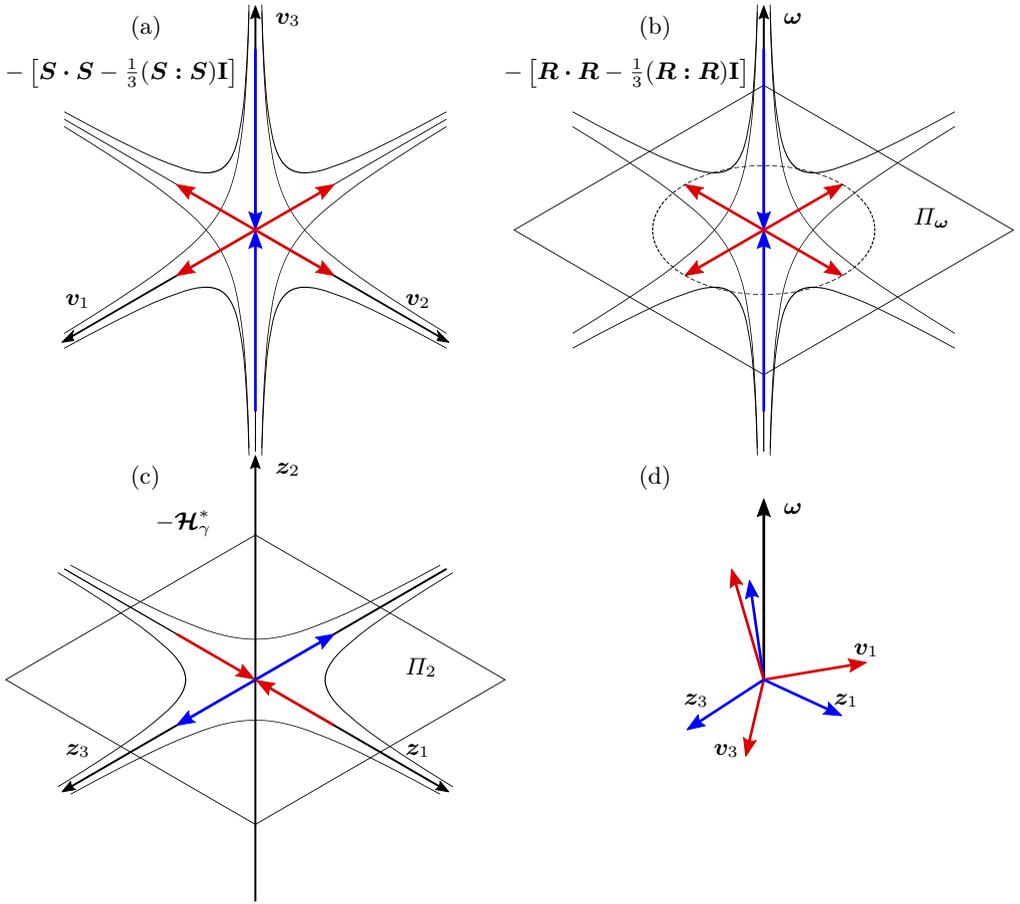

This rank-reduction brings two-dimensional features into three-dimensional flows, and it is interesting to note that the equations for the velocity gradient already contain another two-dimensional flow feature. In particular, in equation \eqref{eq_lambda} the term $(\omega^2 - \widetilde{\omega}_i^2)/4$ arises from the eigenframe representation of $\bm{R\cdot R}=-\omega^2\bm{P_\omega}/4$ where $\bm{P_\omega}$ is the projection tensor on the plane $\Pi_{\bm{\omega}}$ orthogonal to the vorticity vector $\bm{\omega}$. This term describes the straining motion in the plane orthogonal to $\bm{\omega}$ that is associated with the centrifugal force produced by the spinning of the fluid particle about its vorticity axis. As we will discuss later, this two-dimensional effect can be compared with the two-dimensional effect of $\aH_\gamma^*$ on the velocity gradient evolution, leading to interesting insights into their respective dynamical roles. Moreover, $\aH_\gamma^*$ is a two dimensional object in a three-dimensional space which opens the possibility to effectively compare pressure Hessian statistics between two-dimensional and three-dimensional flows. However, the tangent space to the manifold defined by $\aH_\gamma^*$ varies in space and time, therefore the flow on $\Pi_2$ can not be directly compared with Euclidean two-dimensional turbulence but with flows in more complex geometries \citep{Falkovich2014}.

Using the dynamical equaivalence of $\bm{\mathcal{H}}$ and $\bm{\mathcal{H}}_\gamma^*$, we may re-write the equation governing $\lambda_i$ as (ignoring the viscous term)
\begin{equation}
D_t{\lambda_{i}} =  -\left(\lambda^2_{i}-\frac{1}{3}\sum_j\lambda_j^2\right) - \frac{1}{4}\left( \widetilde{\omega}_i^2 - \frac{1}{3}\sum_j\widetilde{\omega}_j^2\right) -\widetilde{\mathcal{H}}^*_{\gamma,i(i)},
\label{eq_lambda_iso}
\end{equation}
and in figure \ref{fig_scheme} we provide a schematic to illustrate the role of each of the terms on the right hand side of \eqref{eq_lambda_iso}.

\section{Numerical results: rank-reduced anisotropic pressure Hessian}
\label{Results}

We now turn to assess the properties of $\aH_\gamma^*$. We do this using data from a Direct Numerical Simulation (DNS) of statistically stationary, isotropic turbulence. The DNS data used are those by \citet{Ireland2016a,Ireland2016b}, at a Taylor microscale Reynolds number $R_\lambda=597$. The data have been obtained through a pseudo-spectral method to solve the incompressible NSE on a three-dimensional, triperiodic cube discretized with $2048^3$ grid points. A deterministic forcing method that preserves the kinetic energy in the flow has been employed. A detailed description of the numerical method used can be found in \citet{Ireland2013}.

\subsection{Pressure Hessian rank reduction}

\begin{figure}
\centering
\begin{overpic}[width=.49\textwidth]{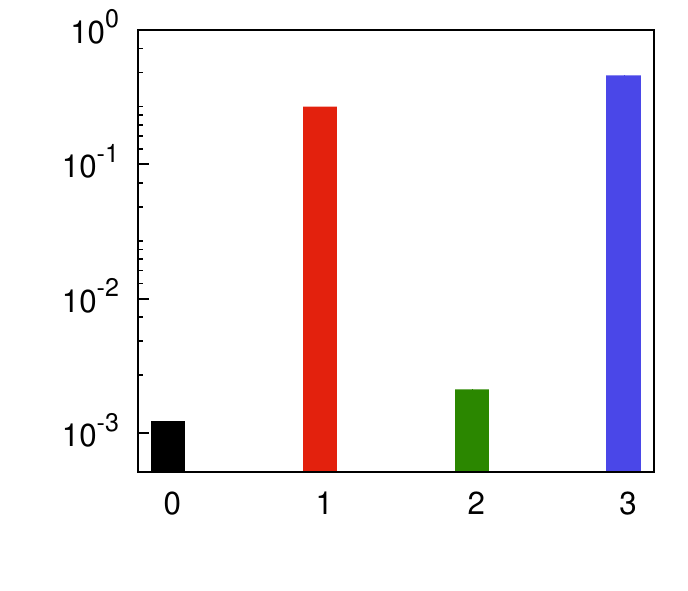}
\put(48,5){{\large $\mu\of{\xi_{RF}}$}}
\put(3,36){\large \rotatebox{90}{Probability}}
\end{overpic}
\caption{Probability of multiplicity of real and finite roots of equation \eqref{eq_psi}.}
\label{fig_pdf_mu_gamma}
\end{figure}

We first consider the properties of $\gamma$ as determined by the numerical solution of equation \eqref{eq_psi} with $\gamma\equiv -1/\xi_{RF}$ real and finite. At each grid point we solve the generalized eigenvalue problem \eqref{eq_gen_evp} to determine real and finite multipliers $\gamma$ for which $\aH_\gamma^*$ is singular.
The numerical solution of equation \eqref{eq_psi} is ill-conditioned when $\det[\bm{R},\bm{S}]$ is very small. Therefore, we skip the grid points at which $\det[\bm{R},\bm{S}]$ is less than a predefined numerical tolerance. We confirmed, however, that the results are only weakly sensitive to this small tolerance value.
Figure~\ref{fig_pdf_mu_gamma} shows the probability of the multiplicity of real and finite values for $\gamma$ obtained solving \eqref{eq_psi}. The statistics are constructed by averaging the flow over space and time, a total of ten snapshots spanning six eddy turnover times have been used.
The rank-reduced anisotropic pressure Hessian exists at the vast majority of the grid points, the configurations with no real and finite multipliers is observed at only about $0.1\%$ of the grid points and corresponds to $\det[\bm{R},\bm{S}]$ very small. The most common case ($\sim60\%$ of the grid points) corresponds to three real and finite roots $\xi_{RF}$ and thus three real and finite multipliers $\gamma$. Therefore, in addition to the continuous symmetry which allows to map $\aH$ into $\aH_\gamma^*$ there is a discrete symmetry, which allows three dynamically equivalent pressure Hessian, which generate the same dynamics of the velocity gradient invariants.
The next most common case ($\sim40\%$ of the grid points) is a single real and finite root $\xi_{RF}$ and so a single $\gamma$ and a single rank-two $\aH_\gamma^*$.
The case with two real and finite roots (and the third root asymptotically small compared with these) is rare ($\sim0.15\%$ of the grid points) and corresponds to $\det[\bm{R},\bm{S}]$ close to zero.
In the configurations in which there exist multiple $\gamma$'s, the multiplier which gives the highest alignment between the vorticity vector and the intermediate eigenvector of the rank-reduced anisotropic pressure Hessian is selected. Indeed, that preferential alignment is a clear feature of the rank-reduced anisotropic pressure Hessian, as we will see below.

\begin{figure}
\begin{overpic}[width=.49\textwidth]{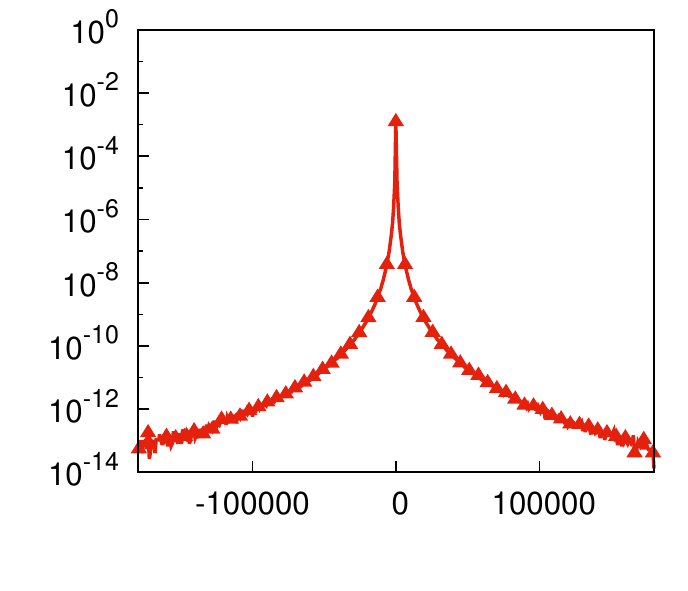}
\put(56,5){{\large $\gamma$}}
\put(2,44){{\large\rotatebox{90}{PDF}}}
\put(22,75) {(a)}
\end{overpic}
\hfill
\begin{overpic}[width=.49\textwidth]{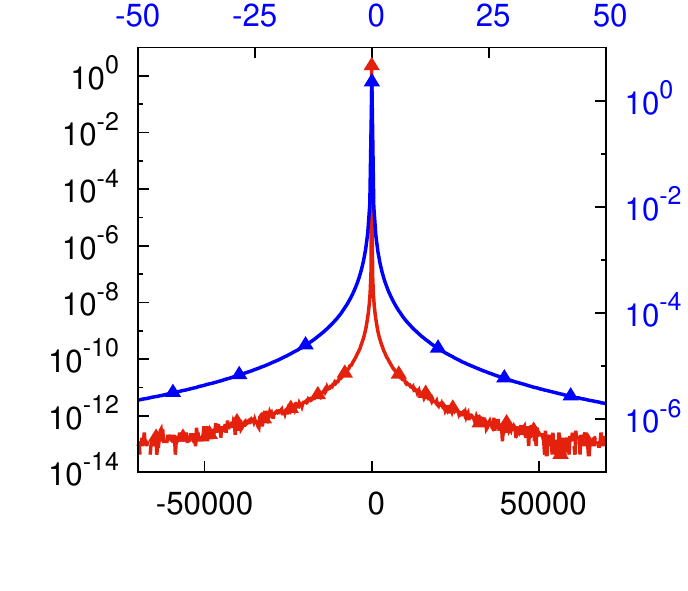}
\put(45,3){\large$\tau_\eta^2\det[\bm{R},\bm{S}]$}
\put(2,44){\large\rotatebox{90}{PDF}}
\put(28,74) {(b)}
\end{overpic}
\begin{overpic}[width=.49\textwidth]{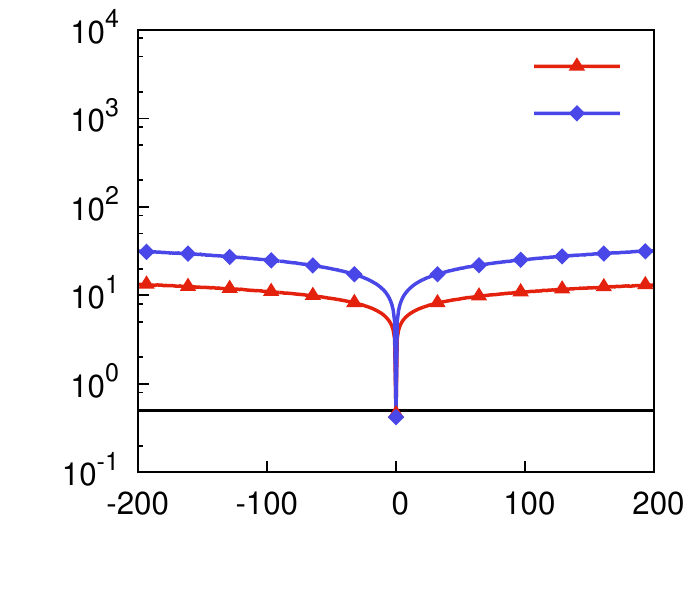}
\put(64,75)  {$\|\bm{S}\|^2$}
\put(64,68.6){$\|\bm{R}\|^2$}
\put(40,75)   {$1/3$}
\put(22,45){\includegraphics[width=.21\textwidth]{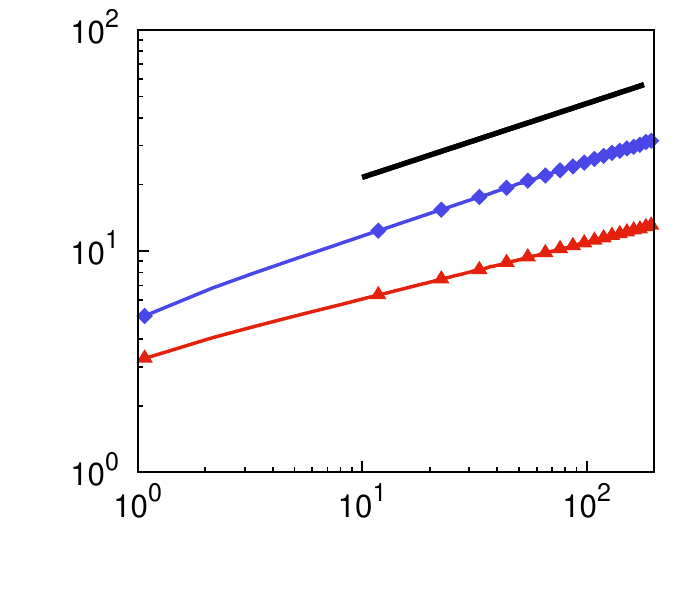}}
\put(46,3)  {\large$\tau_\eta^6\det[\bm{R},\bm{S}]$}
\put(2,30){\rotatebox{90}{\large$\tau_\eta^2\avg{q\big|\det[\bm{R},\bm{S}]}$}}
\put(22,75) {(c)}
\end{overpic}
\hfill
\begin{overpic}[width=.49\textwidth]{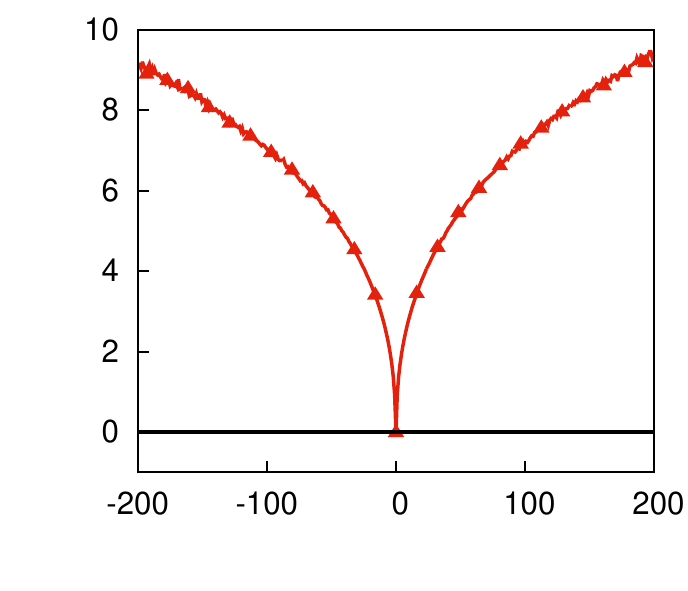}
\put(46,3)  {\large$\tau_\eta^6\det[\bm{R},\bm{S}]$}
\put(2,30){\rotatebox{90}{\large$\tau_\eta^2\avg{Q\big|\det[\bm{R},\bm{S}]}$}}
\put(30,76) {(d)}
\end{overpic}
\caption{(a) Probability density function (PDF) of the real and finite multiplier $\gamma=-1/\xi_{RF}$. (b) PDF of the determinant of the commutator of anti-symmetric and symmetric part of the velocity gradient, $\det[\bm{R},\bm{S}]$,
the blue curve refers to the blue labels and represents the same PDF over a smaller range.
(c) Strain magnitude $\|\bm{S}\|^2$ and rotation magnitude $\|\bm{R}\|^2$ conditioned on $\det[\bm{R},\bm{S}]$, the same plot in logarithmic scale is in the inset.
(d) Second invariant of the velocity gradient tensor $Q$ conditioned on $\det[\bm{R},\bm{S}]$.
}
\label{fig_pdf_gamma}
\end{figure}

The probability density function (PDF) of the multiplier $\gamma$ for which $\aH_\gamma$ has rank two is shown in figure~\ref{fig_pdf_gamma}(a). The PDF of the multiplier is highly non-Gaussian and the multiplier can be very large, even if with a small probability. This is due to the intermittency of the velocity field, that is, large values of the coefficients of equation \eqref{eq_psi} and also due to the high probability of small $\det[\bm{R},\bm{S}]$. In that case indeed, the matrix used for the reduction, $[\bm{R},\bm{S}]$, spans the whole three-dimensional domain but with a very small eigenvalue in a certain eigendirection. As a consequence, the multiplier $\gamma$ should be large enough to compensate the component of $\aH$ in that eigendirection, which can have large values. The probability density function of $\det[\bm{R},\bm{S}]$ is shown in figure \ref{fig_pdf_gamma}(b). The results show that $\det[\bm{R},\bm{S}]$ is highly intermittent, being small throughout the vast majority of the flow, but exhibiting extreme fluctuations is very small regions. This can be understood in terms of the fact that according to equation \eqref{eq_detC}, $\det[\bm{R},\bm{S}]$ is an high-order moment of the velocity gradient field. Moreover, the tendency for small values of $\det[\bm{R},\bm{S}]$ can also be understood in terms of the well-known fact that $\bm{\omega}$ tends to misalign with $\bm{v}_3$ \citep{Meneveau2011}, leading to small values for $\widetilde{\omega}_3$ and therefore to small values of $\det \left[\bm{R},\bm{S}\right ]$ via equation \eqref{eq_detC}.

We now turn to investigate the flow features conditioned on $\det \left[\bm{R},\bm{S}\right ]$.
The high probability to observe small $\det \left[\bm{R},\bm{S}\right ]$ is consistent with the average of the strain and rotation magnitude conditioned on the local value of $\det \left[\bm{R},\bm{S}\right ]$, the results for which are shown in figure \ref{fig_pdf_gamma}(c).
The values of $\tau_\eta^2\|\bm{S}\|^2$ and $\tau_\eta^2\|\bm{R}\|^2$ when $\det[\bm{R},\bm{S}]\to 0$, where $\tau_\eta$ is the Kolmogorov timescale, are both slightly less than $1/2$, that is the precise value of the unconditioned averages $\tau_\eta^2\langle\|\bm{S}\|^2\rangle=\tau_\eta^2\langle\|\bm{R}\|^2\rangle$ in isotropic turbulence. For larger values of $\det \left[\bm{R},\bm{S}\right ]$, $\|\bm{R}\|^2$ has a well defined power law scaling, $\|\bm{R}\|^2\sim\left|\det[\bm{R},\bm{S}]\right|^{1/3}$, as shown in the inset of figure \ref{fig_pdf_gamma}(c). The power law exponent is consistent with simple dimensional analysis. On the other hand, while $\|\bm{S}\|^2$ also depends on $\det[\bm{R},\bm{S}]$ as a power law, the exponent is less than $1/3$, and cannot be predicted by simple dimensional analysis. This is somewhat reminiscent of the results in \cite{Buaria2019} for $\langle \|\bm{R}\|^2\big\vert\|\bm{S}\|^2\rangle$ and $\langle \|\bm{S}\|^2\big\vert\|\bm{R}\|^2\rangle$, where they found that the former was well described by dimensional analysis (i.e.\ by Kolmogorov's 1941 theory, see \cite{pope}), while the latter was not.
The average of the second invariant of the velocity gradient tensor $Q$ conditioned on the local value of $\det[\bm{R},\bm{S}]$ is shown in figure~\ref{fig_pdf_gamma}(d). Interestingly, the region where $\det[\bm{R},\bm{S}]$ is small is slightly strain dominated (i.e.\ $Q<0$). On the other hand, the regions where $|\det[\bm{R},\bm{S}]|$ is relatively large, the dynamics is clearly rotation-dominated. When the conditioned average of $Q$ is weighted with the PDF of $\det[\bm{R},\bm{S}]$ it yields $\avg{Q}=0$ for isotropic turbulence, which indicates the very large relative weight of regions of the flow contributing to $\langle Q\big\vert \det[\bm{R},\bm{S}]\rangle$ being negative and very small.

%

\subsection{Rank-reduced anisotropic pressure Hessian eigenvalue}

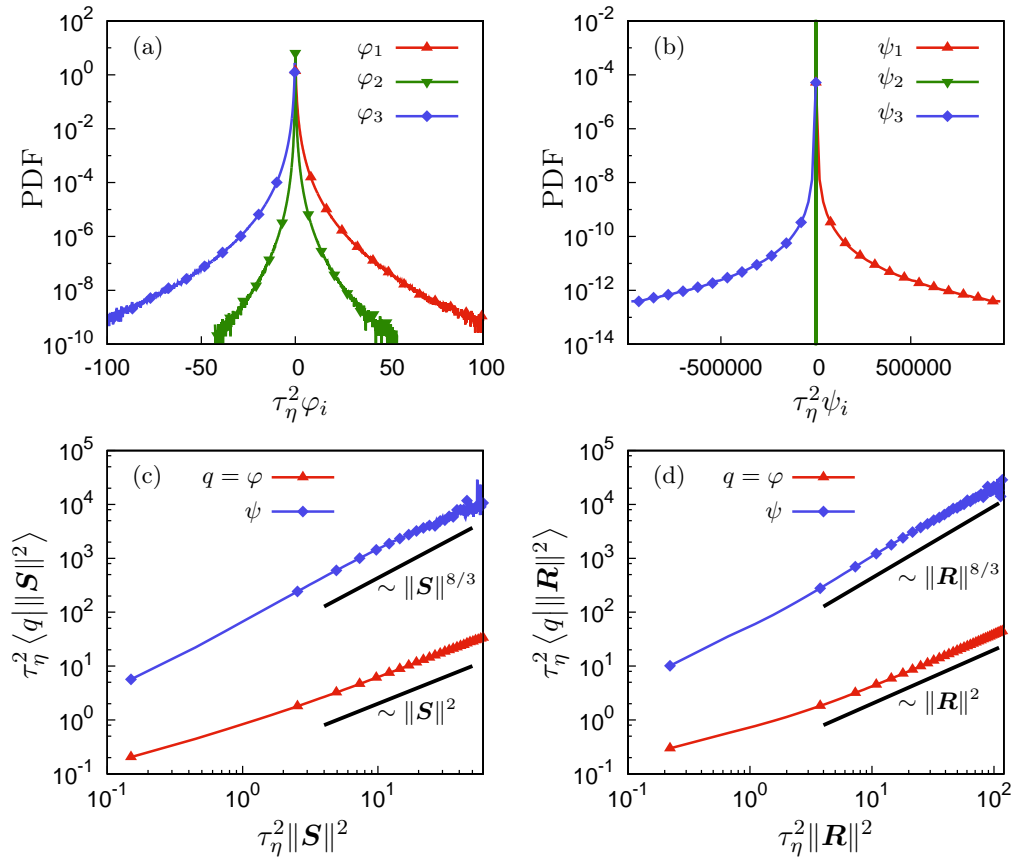
\begin{figure}
\begin{overpic}[width=.49\textwidth]{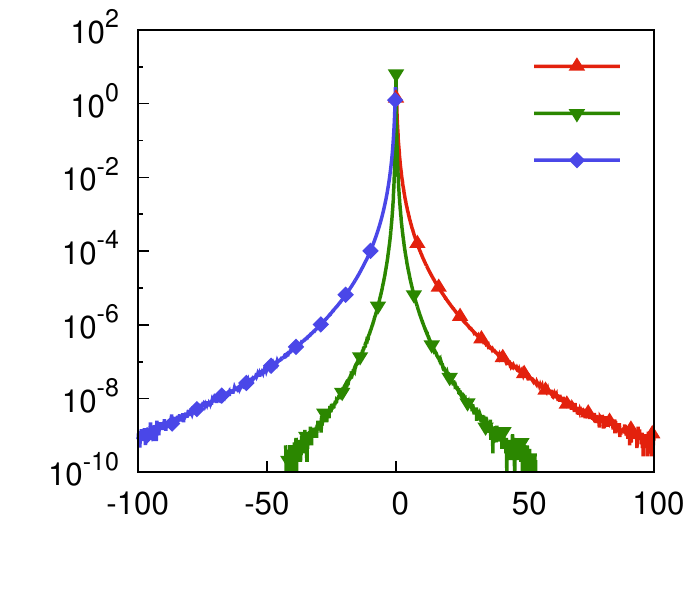}
\put(70,75){$\varphi_1$}
\put(70,69){$\varphi_2$}
\put(70,62){$\varphi_3$}
\put(53,4) {\large$\tau_\eta^2\varphi_i$}
\put(3,44) {\large\rotatebox{90}{PDF}}
\put(25,75) {(a)}
\end{overpic}
\hfill
\begin{overpic}[width=.49\textwidth]{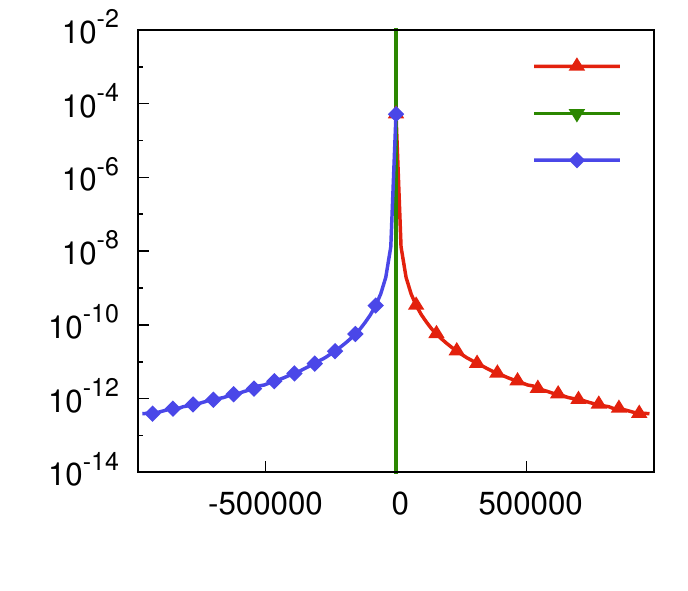}
\put(70,75){$\psi_1$}
\put(70,69){$\psi_2$}
\put(70,62){$\psi_3$}
\put(53,4) {\large$\tau_\eta^2\psi_i$}
\put(3,44) {\large\rotatebox{90}{PDF}}
\put(25,75) {(b)}
\end{overpic}
\vfill
\begin{overpic}[width=.49\textwidth]{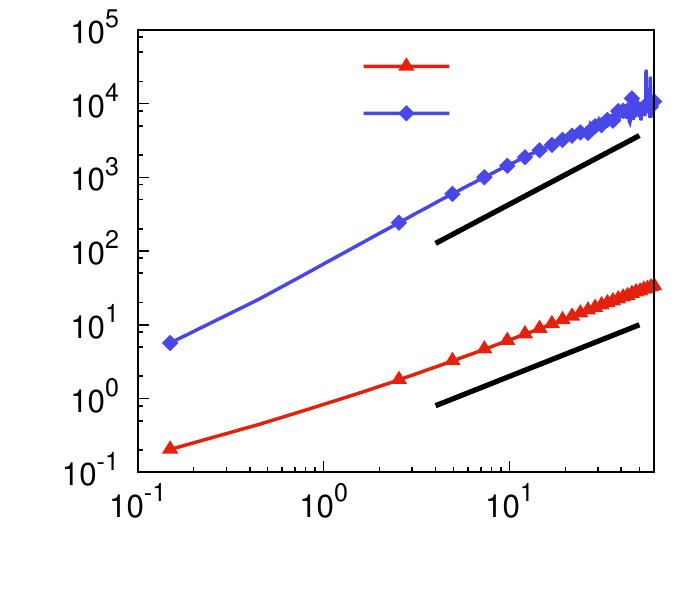}
\put(39,75){$q=\varphi$}
\put(47.5,68){$\psi$}
\put(74,28){$\sim\|\bm{S}\|^2$}
\put(74,53){$\sim\|\bm{S}\|^{8/3}$}
\put(50,3) {\large$\tau_\eta^2\|\bm{S}\|^2$}
\put(1,36) {\large\rotatebox{90}{$\tau_\eta^2\avg{q \big| \|\bm{S}\|^2}$}}
\put(25,75) {(c)}
\end{overpic}
\hfill
\begin{overpic}[width=.49\textwidth]{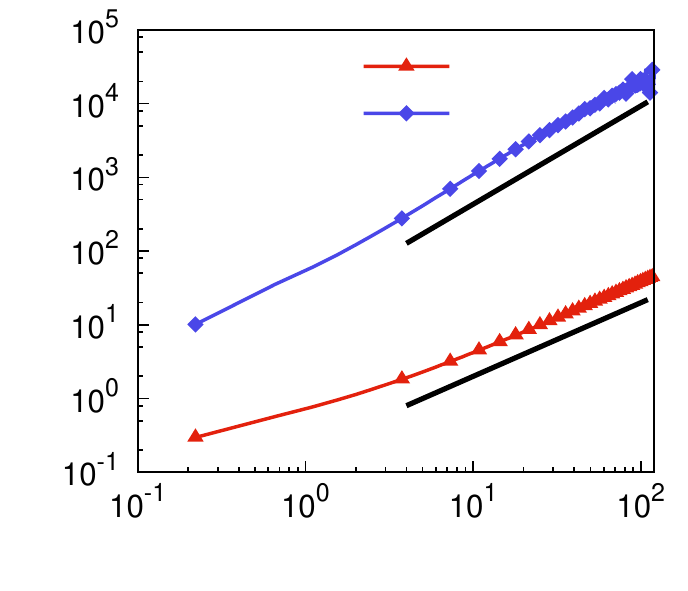}
\put(39,75){$q=\varphi$}
\put(47.5,68){$\psi$}
\put(74,30){$\sim\|\bm{R}\|^2$}
\put(74,55){$\sim\|\bm{R}\|^{8/3}$}
\put(50,3) {\large$\tau_\eta^2\|\bm{R}\|^2$}
\put(1,36) {\large\rotatebox{90}{$\tau_\eta^2\avg{q \big| \|\bm{R}\|^2}$}}
\put(25,75) {(d)}
\end{overpic}
\caption{Probability density function of the eigenvalues of (a) $\aH$ and (b) eigenvalues of $\aH_\gamma^*$, normalized with the Kolmogorov timescale $\tau_\eta$. (c) Magnitude of the anisotropic pressure Hessian eigenvalues $\varphi=\sqrt{\sum_i\varphi_i^2}$ and anisotropic pressure Hessian eigenvalue, $\psi$, conditioned on the local strain-rate magnitude and (d) on the rotation-rate magnitude.}
\label{Hess_eigenvalues}
\end{figure}

The rank-reduction of the anisotropic pressure Hessian corresponds to set its intermediate eigenvalue to zero by means of the gauge term $\gamma[\bm{R},\bm{S}]$. Since the anisotropic pressure Hessian is traceless by definition, $\Tr(\aH)=0$, it has in general two non-zero principal invariants. On the other hand, the rank-reduced anisotropic pressure hesssian has only one non-zero principal invariant, that is $\Tr((\aH_\gamma^*)^2)$ since $\det(\aH_\gamma^*)=0$. 

Figures~\ref{Hess_eigenvalues}(a,b) show that whereas $\aH$ is in general a fully three-dimensional object with three non-zero eigenvalues $\varphi_i$ that satisfy $\sum_{i=1}^3\varphi_i=0$, $\aH_\gamma^*$ is a two-dimensional object with only two active eigenvalues that satisfy $\psi_1=-\psi_3=\psi$, the intermediate eigenvalue being identically zero, $\psi_2=0$. Note that here and throughout, all eigenvectors are unitary, and are ordered according to their corresponding eigenvalues, such that $\varphi_1 \ge \varphi_2 \ge \varphi_3$.
The distributions of the eigenvalues $\varphi_1\ge0$ and $\varphi_3\le 0$ of the anisotropic pressure Hessian display marked tails and are almost symmetric with respect to each other. On the contrary, the distribution of $\varphi_2$ has moderate tails and it is positively skewed.
The eigenvalue of the rank-reduced anisotropic pressure Hessian, $\psi$, exhibits very large fluctuations. Its distribution has wide tails which show that $\psi$, even if with small probability, can take extremely large values. This is in part due to the large intermittency of the flow, giving rise to large values of $[\bm{R},\bm{S}]$ and $\gamma$ (although with small probability). Therefore, the geometrical simplification obtained by replacing the three-dimensional $\aH$ with the two-dimensional $\aH_\gamma^*$ also comes with the cost that the eigenvalue of $\aH_\gamma^*$ is far more intermittent than those of $\aH$.

The large values observed for $\psi$ are also closely related to the dimensionality reduction. In order to investigate this point we condition the eigenvalues of $\aH$ and $\aH_\gamma^*$ on the magnitude of the local strain and vorticity $\|\bm{S}\|^2$ and $\|\bm{R}\|^2$. For the anisotropic pressure Hessian we define 
$\varphi = \sqrt{\sum_i\varphi_i^2}$ and compute the conditional averages
$\avg{\varphi \big| \|\bm{S}\|^2}$ and
$\avg{\varphi \big\vert \|\bm{R}\|^2 }$.
Similarly, for the rank-reduced anisotropic pressure Hessian we look at 
$\avg{\psi \big\vert\|\bm{S}\|^2 }$ and $\avg{\psi \big\vert\|\bm{R}\|^2}$.
The results from the DNS are shown in figures~\ref{Hess_eigenvalues}(c,d). The results reveal a simple scaling $\avg{\varphi \big\vert \|\bm{S}\|^2}\sim\|\bm{S}\|^2$, as dimensional analysis suggests. This lends supports to the model in \cite{Wilczek2014}, in which the pressure Hessian is a linear combination of $\bm{S}^2$, $\bm{R}^2$ and $[\bm{R},\bm{S}]$. The scaling $\avg{\varphi \big\vert \|\bm{S}\|^2}\sim\|\bm{S}\|^2$ is evident especially for large values of $\|\bm{S}\|^2$. This may reflect the idea that during large fluctuations, the lengthscale associated with $\bm{S}$ is smaller as compared to situations where $\bm{S}$ is small or moderate. If true, then the pressure Hessian is more localized during large fluctuations, giving rise to the scaling $\avg{\varphi \big\vert \|\bm{S}\|^2}\sim\|\bm{S}\|^2$ that reflects a local relationship between $\varphi$ and $\|\bm{S}\|^2$. On the other hand, for the rank-reduced anisotropic pressure Hessian eigenvalue we find $\avg{\psi\big\vert\|\bm{S}\|^2}\sim\|\bm{S}\|^{2\zeta}$ with $\zeta>1$ (in particular $\zeta$ between $4/3$ and $5/4$). Nevertheless, $\avg{\psi\big\vert\|\bm{S}\|^2}$ mantains a well defined power law trend, which has positive implications for modelling the anisotropic pressure Hessian using information inferred by the rank-reduced anisotropic pressure Hessian.
Due to the higher exponent, $\psi$ is on average much larger than $\varphi$ at fixed velocity gradient magnitude, especially when large gradients occur.
The scaling of the eigenvalues magnitude conditioned on $\|\bm{R}\|^2$ is very similar to the scaling of the same quantity conditioned on $\|\bm{S}\|^2$ for both $\aH$ and $\aH_\gamma^*$.
The different scaling of $\psi$ and $\varphi$ with respect to the velocity gradient magnitude can be deduced from equation \eqref{eq_rel_psi1}. Indeed, the denominator in equation \eqref{eq_rel_psi1} can be very small since the vorticity tends to align with $\bm{z}_2$, which, as we will see in the next section, inducing large values of $\psi$. This is due to the constraint $\bm{\omega^\top\cdot\aH\cdot\omega}=\bm{\omega^\top\cdot\aH}_\gamma^*\bm{\cdot\omega}$.
From the viewpoint of dimensionality, the rank-reduced anisotropic pressure Hessian is a two-dimensional tensor which has to produce the same effect on the velocity gradient invariants as a fully three-dimensional tensor due to the gauge symmetry. Therefore the geometrical scaling of $\aH_\gamma^*$ is likely to differ from the scaling of $\aH$, which can span the whole three-dimensional embedding space.

\begin{figure}
\includegraphics[width=1.\textwidth]{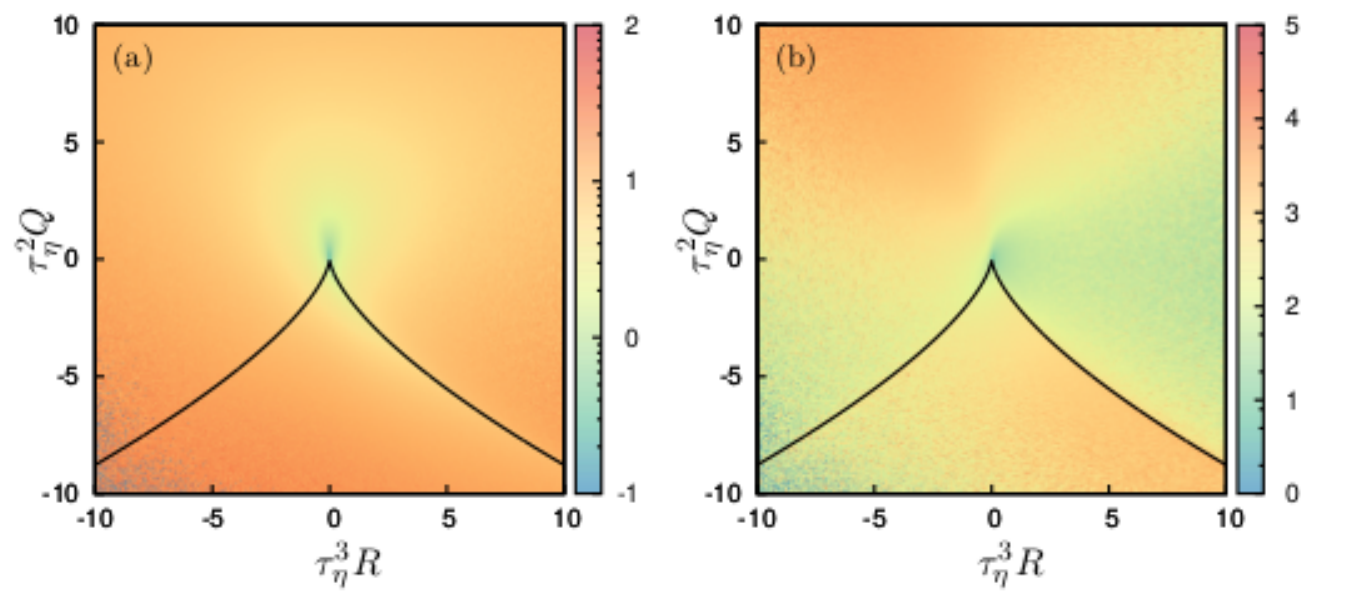}
\caption{Results for (a) $\langle\varphi\big\vert R,Q\rangle$, where $\varphi=\sqrt{\sum_i\varphi_i^2}$, and (b) $\langle\psi\big\vert R,Q\rangle$ as functions of $R,Q$. Colors denote the magnitude of the terms, and black lines denote the Vieillefosse
tails.
}
\label{eigen_QR}
\end{figure}

In figure \ref{eigen_QR} we plot the conditioned averages $\langle\varphi\big\vert R,Q\rangle$ and $\langle\psi\big\vert R,Q\rangle$. The results show that $\langle\varphi\big\vert R,Q\rangle$ is quite large everywhere except for small $R,Q$ and its shape shares similarities with the sheared drop shape of the joint PDF of the invariants $R,Q$, that is in figure \ref{fig_vorticity_S_alignment_RQ}(d). In contrast, $\langle\psi\big\vert R,Q\rangle$ is largest in the quadrants $Q>0,R<0$ and $Q<0,R>0$ (especially below the right Vieillefosse tail) corresponding to regions of enstrophy and strain production. Therefore, it is not only that the magnitudes of $\bm{\mathcal{H}}$ and $\aH_\gamma^*$ differ significantly, but also that they are most active in different regions of the flow. Indeed, $\aH_\gamma^*$ is most active in the regions where the velocity gradients are also most active, while  $\bm{\mathcal{H}}$ is active and strong in many regions where the velocity gradients display relatively little activity (e.g.\ the quadrant $Q<0$, $R<0$). In this sense then, one might say that $\aH_\gamma^*$ is more closely tied to the dynamics of the velocity gradients than $\bm{\mathcal{H}}$.
 
\section{Numerical results: statistical geometry}

\begin{figure}
\begin{overpic}[width=.49\textwidth]{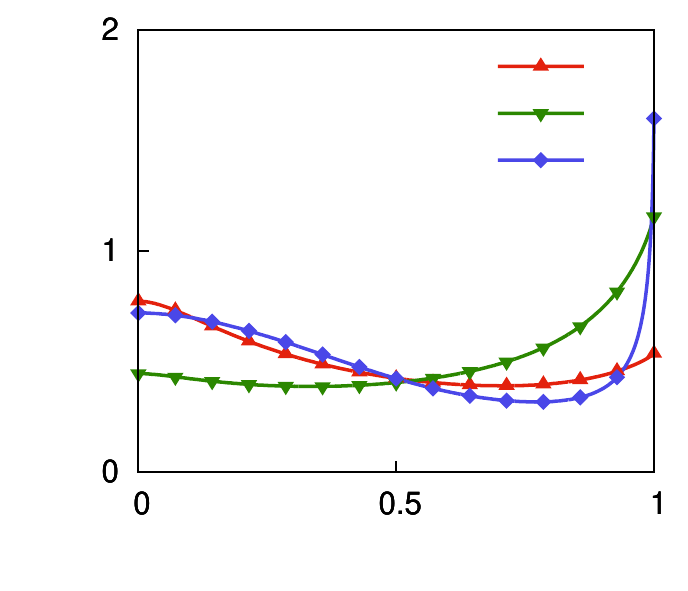}
\put(56,75)  {$\bm{ \hat{\omega} \cdot y}_1$}
\put(56,68.6){$\bm{ \hat{\omega} \cdot y}_2$}
\put(56,61.8){$\bm{ \hat{\omega} \cdot y}_3$}
\put(52,3){\large$\cos\theta$}
\put(3,45){\rotatebox{90}{\large PDF}}
\put(25,75) {(a)}
\end{overpic}
\hfill
\begin{overpic}[width=.49\textwidth]{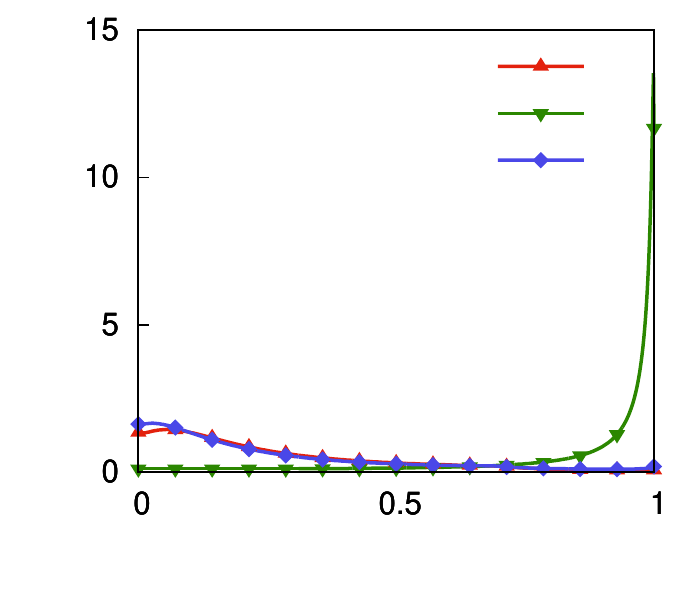}
\put(56,75)  {$\bm{ \hat{\omega} \cdot z}_1$}
\put(56,68.6){$\bm{ \hat{\omega} \cdot z}_2$}
\put(56,61.8){$\bm{ \hat{\omega} \cdot z}_3$}
\put(52,3){\large$\cos\theta$}
\put(3,45){\rotatebox{90}{\large PDF}}
\put(25,75) {(b)}
\end{overpic}
\caption{PDF of the orientation between the vorticity vector and (a) the eigenframe of the pressure Hessian, (b) the eigenframe of the rank-reduced anisotropic pressure Hessian. The alignment is expressed by inner product between the normalized vorticity $\hat{\bm{\omega}}\equiv\bm{\omega}/\|\bm{\omega}\|$ and normalized eigenvetors of $\aH$ ($\bm{y}_i$) and $\aH_\gamma^*$ ($\bm{z}_i$).}
\label{fig_omgH_align}
\end{figure}

\begin{figure}
\begin{overpic}[width=.49\textwidth]{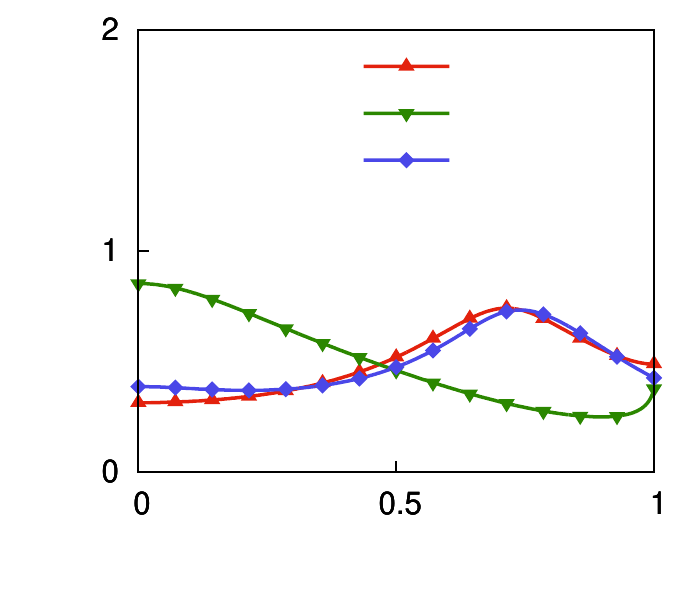}
\put(36,75)  {$\bm{y}_1\bm{\cdot}\bm{v}_1$}
\put(36,68.6){$\bm{y}_1\bm{\cdot}\bm{v}_2$}
\put(36,61.8){$\bm{y}_1\bm{\cdot}\bm{v}_3$}
\put(52,3){\large$\cos\theta$}
\put(3,45){\large\rotatebox{90}{PDF}}
\put(25,75) {(a)}
\end{overpic}
\hfill
\begin{overpic}[width=.49\textwidth]{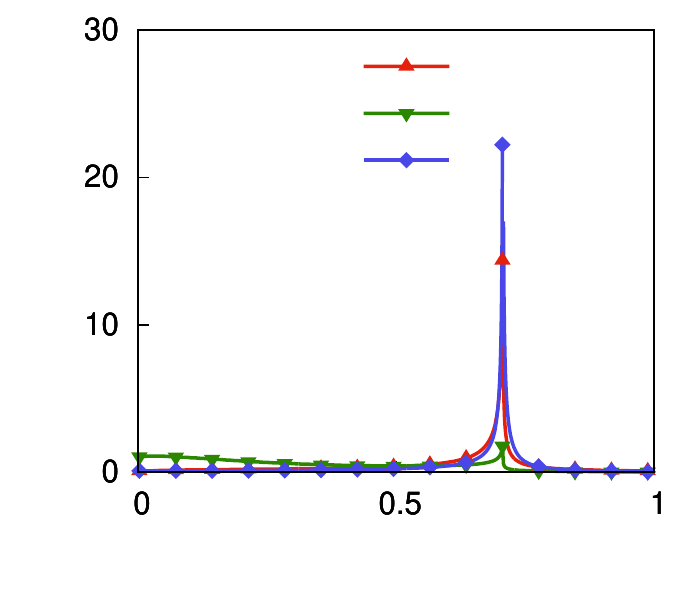}
\put(36,75)  {$\bm{z}_1\bm{\cdot}\bm{v}_1$}
\put(36,68.6){$\bm{z}_1\bm{\cdot}\bm{v}_2$}
\put(36,61.8){$\bm{z}_1\bm{\cdot}\bm{v}_3$}
\put(52,3){\large$\cos\theta$}
\put(3,45){\large\rotatebox{90}{PDF}}
\put(25,75) {(b)}
\end{overpic}
\vfill
\begin{overpic}[width=.49\textwidth]{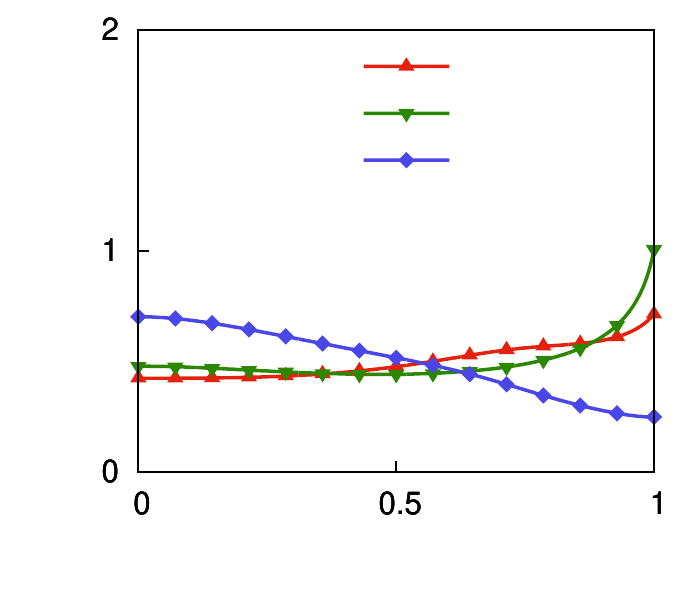}
\put(36,75)  {$\bm{y}_2\bm{\cdot}\bm{v}_1$}
\put(36,68.6){$\bm{y}_2\bm{\cdot}\bm{v}_2$}
\put(36,61.8){$\bm{y}_2\bm{\cdot}\bm{v}_3$}
\put(52,3){\large$\cos\theta$}
\put(3,45){\large\rotatebox{90}{PDF}}
\put(25,75) {(c)}
\end{overpic}
\hfill
\begin{overpic}[width=.49\textwidth]{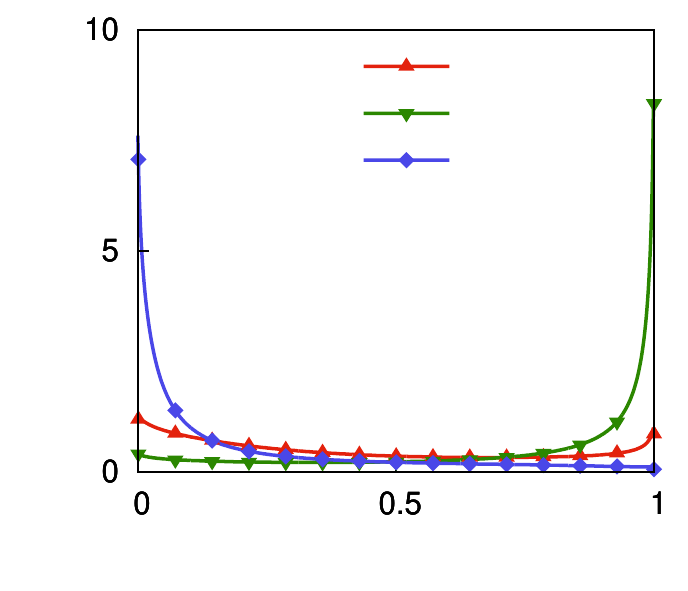}
\put(36,75)  {$\bm{z}_2\bm{\cdot}\bm{v}_1$}
\put(36,68.6){$\bm{z}_2\bm{\cdot}\bm{v}_2$}
\put(36,61.8){$\bm{z}_2\bm{\cdot}\bm{v}_3$}
\put(52,3){\large$\cos\theta$}
\put(3,45){\large\rotatebox{90}{PDF}}
\put(25,75) {(d)}
\end{overpic}
\vfill
\begin{overpic}[width=.49\textwidth]{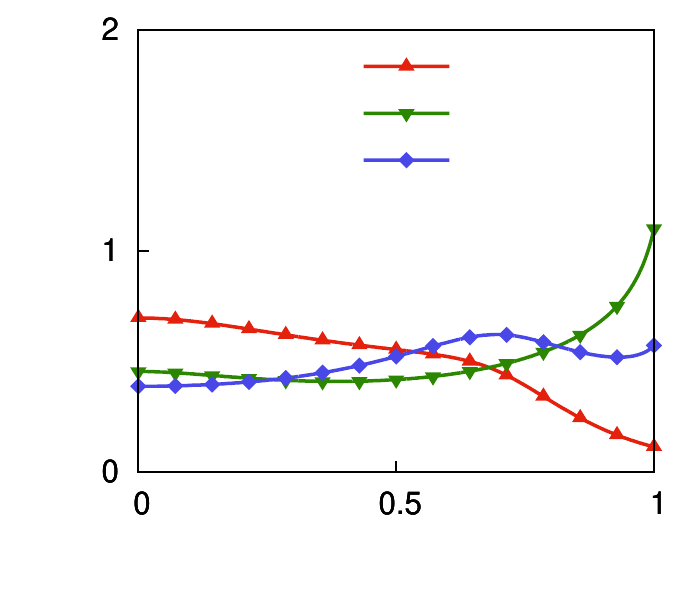}
\put(36,75)  {$\bm{y}_3\bm{\cdot}\bm{v}_1$}
\put(36,68.6){$\bm{y}_3\bm{\cdot}\bm{v}_2$}
\put(36,61.8){$\bm{y}_3\bm{\cdot}\bm{v}_3$}
\put(52,3){\large$\cos\theta$}
\put(3,45){\large\rotatebox{90}{PDF}}
\put(25,75) {(e)}
\end{overpic}
\hfill
\begin{overpic}[width=.49\textwidth]{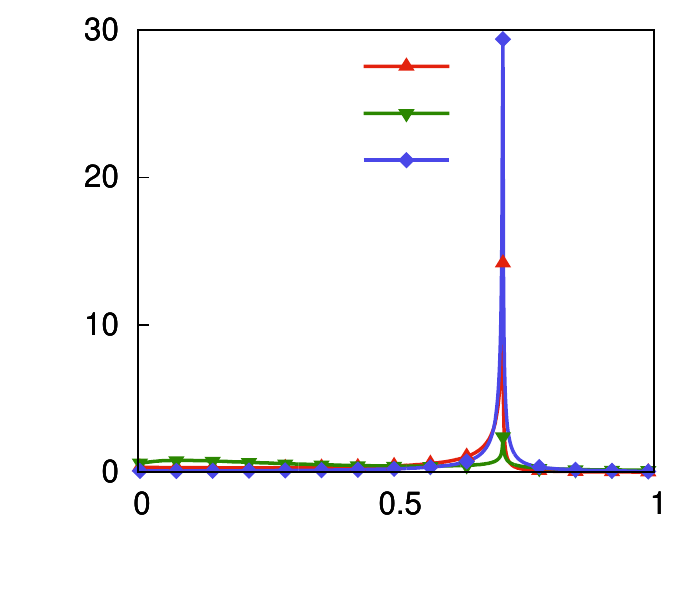}
\put(36,75)  {$\bm{z}_3\bm{\cdot}\bm{v}_1$}
\put(36,68.6){$\bm{z}_3\bm{\cdot}\bm{v}_2$}
\put(36,61.8){$\bm{z}_3\bm{\cdot}\bm{v}_3$}
\put(52,3){\large$\cos\theta$}
\put(3,45){\large\rotatebox{90}{PDF}}
\put(25,75) {(f)}
\end{overpic}
\caption{PDF of the relative orientation between the pressure Hessian eigenframe and the strain eigenframe (a-c-e) and relative orientation between the rank-reduced anisotropic pressure Hessian eigenframe and the strain eigenframe (b-d-f). The orientation is expressed by inner product of the eigenvectors of the strain-rate tensor $\bm{v}_i$ with the eigenvectors of $\aH$ ($\bm{y}_i$) and the eigenvectors of $\aH_\gamma^*$   ($\bm{z}_i$).}
\label{fig_HS_align}
\end{figure}

We now turn to consider the statistical geometry of the system.
In figure~\ref{fig_omgH_align} we consider the alignment between the vorticity $\bm{\omega}$ and the eigenframes of $\aH$ and $\aH_\gamma^*$.
While there is a strong preferential statistical alignment of the intermediate strain-rate eigenvector $\bm{v}_2$ with $\bm{\omega}$ \citep{Meneveau2011}, the  preferential statistical alignment between $\bm{\omega}$ and the pressure Hessian eigenvectors $\bm{y}_i$ is very weak. There is only a moderate tendency for alignment between $\bm{y}_{2,3}$ and $\bm{\omega}$ \citep{Chevillard2008}. This constitutes an obstacle for understanding the role of the anisotropic pressure Hessian in turbulence.

On the other hand, the results in figure~\ref{fig_omgH_align} show a striking alignment between $\bm{\omega}$ and the rank-reduced anisotropic pressure Hessian eigenvectors $\bm{z}_i$. Indeed, there is a remarkable tendency for $\bm{z}_2$ to align with $\bm{\omega}$, that is consistent with the preferential alignment between $\bm{v}_2$ and $\bm{\omega}$ and between $\bm{z}_2$ and $\bm{v}_2$ (figure \ref{fig_HS_align}). As discussed in \S\ref{RR}, the contribution of the vorticity and rank-reduced anisotropic pressure Hessian to the straining motion in the fluid is confined to planes. In particular, the straining associated with the centrifugal force produced by the spinning of the fluid particle about the vorticity axis acts in the plane $\Pi_{\bm{\omega}}$, orthogonal to $\bm{\omega}$, while the contribution from the rank-reduced anisotropic pressure Hessian lies on the plane $\Pi_2$, orthogonal to its intermediate eigenvector $\bm{z}_2$. The results shown in figure \ref{fig_omgH_align}(b) indicate that these two planes tend to almost coincide.
However, the effects of $\bm{\omega}$ and $\aH_\gamma^*$ on the strain-rate dynamics are radically different.
The rotation of the fluid element generates a stretching rate of magnitude $\omega^2/4$ on the plane $\Pi_{\bm{\omega}}$ and its contribution is isotropic, since the eigenvalue of the projection tensor $\bm{P_\omega}$ is the same for all the eigenvectors that belong to the plane $\Pi_{\bm{\omega}}$, as in figure \ref{fig_scheme}(b).
On the other hand, the rank-reduced anisotropic pressure Hessian causes a stretching rate of magnitude $\psi$ in direction $\bm{z}_3$ and an equal and opposite compression in the direction $\bm{z}_1$, orthogonal to $\bm{z}_3$, as in figure \ref{fig_scheme}(c). This results in a marked anisotropy of the effect of $\aH_\gamma^*$ on the plane $\Pi_2$.
Since the planes $\Pi_{\bm{\omega}}$ and $\Pi_2$ tend to be almost parallel, the anisotropic pressure Hessian can be understood as the cause of the anisotropy which is lacking in the centrifugal forces produced by the vorticity in the $\Pi_{\bm{\omega}}$ plane, and this anisotropy is a key element in the prevention of the blow-up of the system.

Interestingly, the gauge term used in defining $\aH_\gamma^*$, equation \eqref{eq_H_gamma}, arises from a rotation of the strain-rate eigenframe about $\bm{\omega}$ and the results show that $\aH_\gamma^*$ lives on a two-dimensional manifold that statistically has a strong, but imperfect tendency to be orthogonal to $\bm{\omega}$. The dynamical significance of the slight misalignment is that it allows the anisotropic pressure Hessian to contribute to the eigenframe dynamics.
Indeed, if $\aH_\gamma^*$ were exactly orthogonal to $\bm{\omega}$, then the anisotropic pressure Hessian would make no direct contribution to the vorticity dynamics, and its only role would be to contribute to the strain-rate dynamics, described by equation \eqref{eq_lambda} and \eqref{eq_alg}.
It is known that in the inviscid case, the neglect of the anisotropic pressure Hessian in the eigenframe dynamics leads to a finite time singularity \citep{Vieillefosse1982}. Therefore, assuming that the slight misalignment between $\aH_\gamma^*$ and $\bm{\omega}$ is not solely due to viscous effects, then this misalignment must also play a role in regularizing the eigenframe dynamics
thus preventing the onset of singularities in the inviscid Euler system.

Figures~\ref{fig_HS_align}(a-c-e) present the statistical alignment of the eigenvectors $\bm{y}_i$ of $\aH$, with the strain-rate eigenvectors $\bm{v}_j$. The alignments between the pressure Hessian eigenframe and the strain-rate eigenframe do not reveal any strong preferences, with weak alignment tendencies to $\bm{y}_1\bm{\cdot}\bm{v}_1\approx 0.71$ and $\bm{y}_{1,3}\bm{\cdot}\bm{v}_3\approx 0.71$. Therefore, there is a very mild tendency for $\bm{y}_1$ to form a $\pi/4$ angle with $\bm{v}_1$ and $\bm{v}_3$ and for $\bm{y}_3$ to form a $\pi/4$ angle with $\bm{v}_3$. These weak alignments make it difficult to model the directionality of $\aH$ in any simple way in terms of the eigenframe of the strain-rate tensor.

Figure~\ref{fig_HS_align}(b-d-f) show the alignments between the eigenvectors $\bm{z}_i$ of $\aH_\gamma^*$, with $\bm{v}_j$. The results show, in striking contrast to the corresponding plots for the alignment of $\aH$, that the eigenframe $\aH_\gamma^*$ exhibits remarkable alignment properties with a strong tendency to have $\bm{z}_{1,3}\bm{\cdot}\bm{v}_{1,3}\approx 0.71$, $\bm{z}_{2}\bm{\cdot}\bm{v}_2\approx 1$ and $\bm{z}_{2}\bm{\cdot}\bm{v}_3\approx 0$. This means that the tangent space $\Pi_2$ to the two-dimensional manifold on which $\aH_\gamma^*$ acts tends to be orthogonal to $\bm{v}_2$. On that plane the eigenvetors $\bm{z}_1$ and $\bm{z}_3$ of $\aH_\gamma^*$ tend to be inclined at an angle of $\pi/4$ relative to both $\bm{v}_1$ and $\bm{v}_3$. This evidence makes the rank-reduced anisotropic pressure Hessian suitable for modelling, since there is a well defined most probable configuration for the orientation of $\aH_\gamma^*$ with respect to $\bm{S}$.
Those clear preferential alignments between $\bm{\omega}$ and $\bm{S}$ with $\aH_\gamma^*$ also helps understanding how the anisotropic pressure Hessian prevents blow-up, as we will discuss in the next section.

\section{Numerical results: conditioned statistical geometry}

The simpler geometry of the rank-reduced anisotropic pressure Hessian together with its well-defined preferential alignments can facilitate the understanding of the pressure Hessian on the dynamics of the velocity gradient invariants.
In particular, the role of the anisotropic pressure Hessian in preventing the blow-up of the Restricted Euler system can be analyzed by considering how the statistical alignment properties of $\aH_\gamma^*$ depend on $\bm{S}$ and $\bm{\omega}$.

The finite-time singularity prevention mechanism can be safely tackled by using $\aH_\gamma^*$ instead of $\aH$ since such regularity problem is expressed in terms of invariants and is not linked with the orientation of the strain-rate eigenframe with respect to a fixed frame. The equations for the invariants dynamics, \eqref{eq_lambda} and \eqref{eq_omega}, show that there is a local stabilizing effect due to the reduction of the strain-rates by the centrifugal force produced by the vorticity. However, it is known that this mechanism alone is not sufficient to prevent blow-up of the system \citep{Meneveau2011}, and the anisotropic pressure Hessian provides the additional contribution to stabilize the dynamics. This can be understood more easily when the rank-reduced anisotropic pressure Hessian is employed instead of the full anisotropic pressure Hessian.
Indeed, $\aH_\gamma^*$ is effective only on a plane and the results show a clear tendency for $\bm{S}$  and $\bm{\omega}$  to preferentially align with $\aH_\gamma^*$, which is in striking contrast with their mild preferential alignment with $\aH$.

\subsection{Rank-redued anisotropic pressure Hessian--strain-rate alignment}

The  components of the rank-redued anisotropic pressure Hessian, $\aH_\gamma^*$, in the strain-rate eigenframe can be expressed as
\begin{equation}
\bm{V}^\top\bm{\cdot}\aH_\gamma^*\bm{\cdot}\bm{V} = \bm{V}^\top\bm{\cdot} \bm{Z\cdot\psi \cdot Z}^\top \bm{\cdot}\bm{V}
\label{eq_H_components}
\end{equation}
where $\bm{V}$ and $\bm{Z}$ are the matrices which contain the strain-rate eigenvectors  components and rank-reduced pressure Hessian eigenvectors components with respect to a Cartesian basis, that is, $V_{ij}\equiv\bm{e}_i\bm{\cdot v}_j$ and $Z_{ij}\equiv\bm{e}_i\bm{\cdot z}_j$.
The diagonal and singular matrix $\bm{\psi}$ contains the eigenvalues of the rank-reduced anisotropic pressure Hessian, $(\psi,0,-\psi)$.
The components of $\aH_\gamma^*$ in the strain-rate eigenframe can be explicitly computed,
\begin{equation}
\widetilde{\aH}_\gamma^* = \bm{V}^\top\bm{\cdot}\aH_\gamma^*\bm{\cdot}\bm{V} = \psi
\begin{bmatrix}
\widetilde{{z}}_{11}^2 - \widetilde{{z}}_{13}^2 &
\widetilde{{z}}_{11}     \widetilde{{z}}_{21} - \widetilde{{z}}_{13}\widetilde{{z}}_{23} &
\widetilde{{z}}_{11}     \widetilde{{z}}_{31} - \widetilde{{z}}_{13}\widetilde{{z}}_{33} \\
\widetilde{{z}}_{11}     \widetilde{{z}}_{21} - \widetilde{{z}}_{13}\widetilde{{z}}_{23} &
\widetilde{{z}}_{21}^2 - \widetilde{{z}}_{23}^2 &
\widetilde{{z}}_{21}     \widetilde{{z}}_{31} - \widetilde{{z}}_{23}\widetilde{{z}}_{33} \\
\widetilde{{z}}_{11}     \widetilde{{z}}_{31} - \widetilde{{z}}_{13}\widetilde{{z}}_{33} &
\widetilde{{z}}_{21}     \widetilde{{z}}_{31} - \widetilde{{z}}_{23}\widetilde{{z}}_{33} &
\widetilde{{z}}_{31}^2 - \widetilde{{z}}_{33}^2 \\
\end{bmatrix},
\label{eq_H_gamma_gen}
\end{equation}
where $\widetilde{z}_{ij}\equiv\bm{v}_i\bm{\cdot z}_j$ is the $i$-th strain-rate eigenframe component of the $j$-th eigenvector $\bm{z}_j$ and $\sum_i \widetilde{z}_{ij}^2=1$. Since $\aH_\gamma^*$ acts only on the plane $\Pi_2$, spanned by $\bm{z}_1$ and $\bm{z}_3$, the expression of $\aH_\gamma^*$ in the strain-rate eigenframe is simplified. The rank-reduction allows for separation of variables between the magnitude and orientation contributions.
The magnitude of the pressure Hessian is described solely by $\psi$ while the orientation depends on the dot products $\widetilde{z}_{ij}$.
The factorization into the product of a function only of the eigenvalue and a function only of the alignment of the eigenframes is a feature of two-dimensional traceless tensors, while in three dimensions such separation of variables is in general not possible \citep{Ballouz2018}.

\begin{figure}
\includegraphics[width=1.\textwidth]{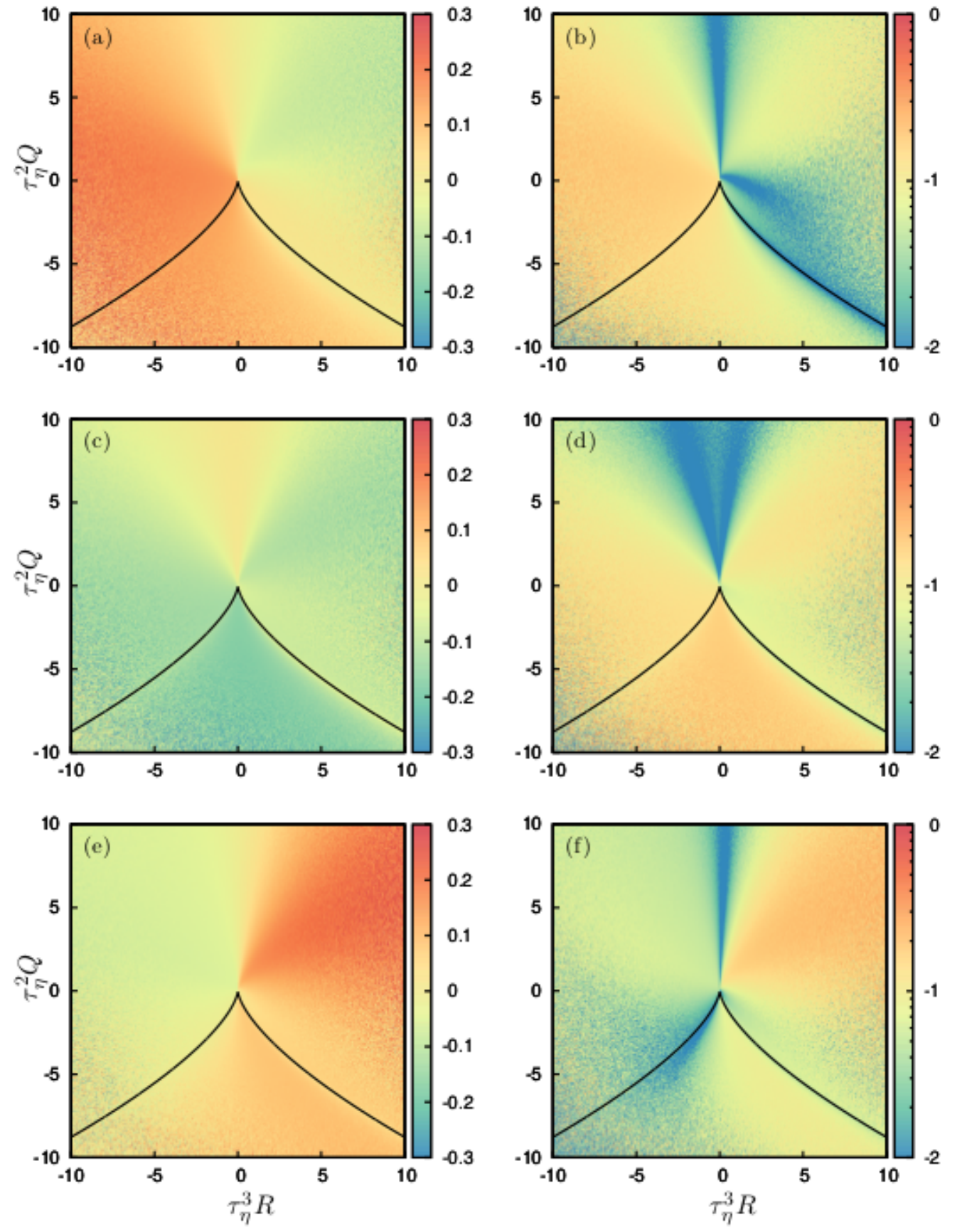}
\caption{Results for $\langle\widetilde{{z}}_{i3}^2 - \widetilde{{z}}_{i1}^2\big\vert R,Q\rangle$, (a) $i=1$, (c) $i=2$, (e) $i=3$. The color range has been truncated to $[-0.3,0.3]$ in order to highlight the trend of the variables around the most probable values.
Results for $\langle|\widetilde{{z}}_{i3}^2 - \widetilde{{z}}_{i1}^2|\big\vert R,Q\rangle$ in logarithmic scale, (b) $i=1$, (d) $i=2$, (f) $i=3$. Black lines denote the Vieillefosse
tails.
}
\label{blow_up}
\end{figure}

The diagonal components of $\aH_\gamma^*$ in the strain-rate eigenframe cause a variation of the strain-rate eigenvalues. Using equation \eqref{eq_H_gamma_gen} in \eqref{eq_lambda}, and neglecting the viscous contribution, gives
\begin{equation}
D_t{\lambda_{i}} =  -\left(\lambda^2_{i}-\frac{1}{3}\sum_j\lambda_j^2\right) - \frac{1}{4}\left( \widetilde{\omega}_i^2 - \frac{1}{3}\sum_j\widetilde{\omega}_j^2\right) - \psi \left(\widetilde{{z}}_{i1}^2 - \widetilde{{z}}_{i3}^2\right).
\label{eq_lambda_z}
\end{equation}
It is known that the blow up of the Restricted Euler model occurs in the quadrant $R>0,Q<0$ where 
the invariants $R$ and $Q$ are defined in equation \eqref{eq_def_RQ}. In particular, the blow-up is associated with  $R\to+\infty$ and $Q\sim-(27R^2/4)^{1/3}\to-\infty$ \citep{Vieillefosse1982}. In this quadrant the straining field is in a state of bi-axial extension, with $\lambda_1>0,\lambda_2>0,\lambda_3<0$. Therefore, to explore how $\aH_\gamma^*$ prevents blow-up, we must consider its effects on the states where $\lambda_1>0,\lambda_2>0,\lambda_3<0$.
From equation \eqref{eq_lambda_z} we see that $\aH_\gamma^*$ will act to prevent blow-up in the quadrant $Q<0, R>0$ if $\widetilde{{z}}_{13}^2 - \widetilde{{z}}_{11}^2<0$, $\widetilde{{z}}_{23}^2 - \widetilde{{z}}_{21}^2<0$, and $\widetilde{{z}}_{33}^2 - \widetilde{{z}}_{31}^2>0$. To consider this, in figure \ref{blow_up} we show the conditioned averages $\langle\widetilde{{z}}_{i3}^2 - \widetilde{{z}}_{i1}^2\big\vert R,Q\rangle$. The results confirm that when $Q<0, R>0$, $\langle\widetilde{{z}}_{23}^2 - \widetilde{{z}}_{21}^2\big\vert R,Q\rangle<0$, and $\langle\widetilde{{z}}_{33}^2 - \widetilde{{z}}_{31}^2\big\vert R,Q\rangle>0$, showing that $\aH_\gamma^*$ acts to reduce $|\lambda_2|$ and $|\lambda_3|$. However, contrary to expectation, they also show that $\langle\widetilde{{z}}_{13}^2 - \widetilde{{z}}_{11}^2\big\vert R,Q\rangle>0$, such that $\aH_\gamma^*$ explicitly acts to increase $\lambda_1$ when $Q<0, R>0$.  Nevertheless, since $\sum_i \lambda_i =0$, if $\aH_\gamma^*$ acts to reduce $|\lambda_3|$ when $Q<0, R>0$, then it also indirectly acts to reduce $\lambda_1$, since $\lambda_1\to\infty$ is not possible unless $|\lambda_3|\to\infty$ (noting $-\lambda_3\geq \lambda_2$). Therefore, the effect of $\aH_\gamma^*$ is somewhat subtle, directly acting to prevent blow-up of $\lambda_2$ and $\lambda_3$, and only indirectly acting to prevent the blow-up of $\lambda_1$. Interestingly, the direct amplification of $\lambda_1$ due to $\aH_\gamma^*$ becomes very small in a narrow region along the right Vieillefosse tail, as the colors in figure \ref{blow_up}(b) show. Therefore, this amplification mechanism is not effective in the phase space region in which the Restricted Euler system blows up.

The scalar products $\widetilde{z}_{ij}$ preferentially lie in a very narrow interval around a few well defined values, as clearly indicated by the results in figure \ref{fig_HS_align}.
In particular, the eigenvectors $\bm{z}_{1,3}$ of $\aH_\gamma^*$ tend to form an angle of $\pi/4$ with the eigenvectors $\bm{v}_{1,3}$ of $\bm{S}$. Therefore a typical configuration for the relative orientation between $\aH_\gamma^*$ and $\bm{S}$ is
\begin{equation}
\bm{V}^\top\bm{\cdot Z} =
\begin{bmatrix}
\cos\left(\pi/4+\epsilon_{11}\right) & \sin\left(\epsilon_{12}\right) &  \cos\left(\pi/4+\epsilon_{13}\right)\\
\sin\left(\epsilon_{21}\right) & \cos\left(\epsilon_{22}\right) & \sin\left(\epsilon_{23}\right)\\
\cos\left(\pi/4+\epsilon_{31}\right) & \sin\left(\epsilon_{32}\right) & \cos\left(\pi/4+\epsilon_{33}\right)\\
\end{bmatrix},
\label{eq_rot_VZ}
\end{equation}
where the quantities $\epsilon_{ij}$ represent the deviations of the angles from the idealized configuration considered, and there is a dependence of the sign on the angle between $\bm{v}_1$ and $\bm{z}_1$, which can be $\pi/4$ or $3\pi/4$ (depending upon the sign of the eigenvalues that are chosen). That sign does not change the discussion below.
Considering only small deviations from the most probable alignment, that is, considering $|\epsilon_{ij}|\ll1$, the elements of the rotation matrix in equation \eqref{eq_rot_VZ} can be Taylor-expanded and, at first order in $\epsilon_{ij}$, the expression for the rank-reduced anisotropic pressure Hessian in the strain-rate eigenframe reduces to
\begin{equation}
\widetilde{\aH}_\gamma^* = \bm{V}^\top\bm{\cdot}\aH_\gamma^*\bm{\cdot V} \sim
\psi
\begin{bmatrix}
-2\epsilon_{11} & \epsilon_{32} & \pm 1 \\
\epsilon_{32} & 0 & \epsilon_{12} \\
 \pm 1 & \epsilon_{12} & 2\epsilon_{11}\\
\end{bmatrix},
\label{eq_H_gamma_V}
\end{equation}
where the orthonormality constraint, $\bm{V\cdot V}^\top=\mathbf{I}$, has been used to relate the small perturbation angles.
It is the diagonal components of $\aH_\gamma^*$ that contribute directly to the rate of change of the strain-rate eigenvalues, as in equation \eqref{eq_lambda_z}, and the anisotropic pressure Hessian has no direct effect on the strain-rate eigenvalues when the most probable alignments, $\epsilon_{ij}=0$, occur.
At the level of this first order approximation, the effect of $\aH_\gamma^*$ on the first and third eigenvalue always has the opposite sign, which is consistent with the stabilizing effect of the pressure Hessian. Therefore, according to this first order approximation, the pressure Hessian tends to counteract both $\lambda_1$ and $\lambda_3$ by imposing a negative rate of change of $\lambda_1$ and a positive rate of change of $\lambda_3$, such that both the most positive and negative eigenvalues are pulled toward smaller magnitudes. The results in figure \ref{blow_up} confirm this prediction in the $Q>0,R>0$ quadrant,
where it is seen that ${\aH}_\gamma^*$ acts to suppress the magnitudes of both $\lambda_1$ and $\lambda_3$. That the linearized prediction fails in the region $Q<0,R>0$ is perhaps not surprising since that is the region of most intense nonlinear activity, and where ${\aH}_\gamma^*$ must be sufficiently large (and by implication $\epsilon_{ij}$ cannot be too small) in order to counteract the blow-up associated with the RE dynamics. The linearization also predicts that the influence of $\aH_\gamma^*$ on $\lambda_2$ is only a second order effect when $\epsilon_{ij}$ is small. However, this prediction is in general not supported by the DNS, since the results in figure \ref{blow_up} show that in most of the $Q,R$ plane, the rank-reduced anisotropic pressure Hessian strongly hinders the growth of positive $\lambda_2$. 

\begin{figure}
\includegraphics[width=1.\textwidth]{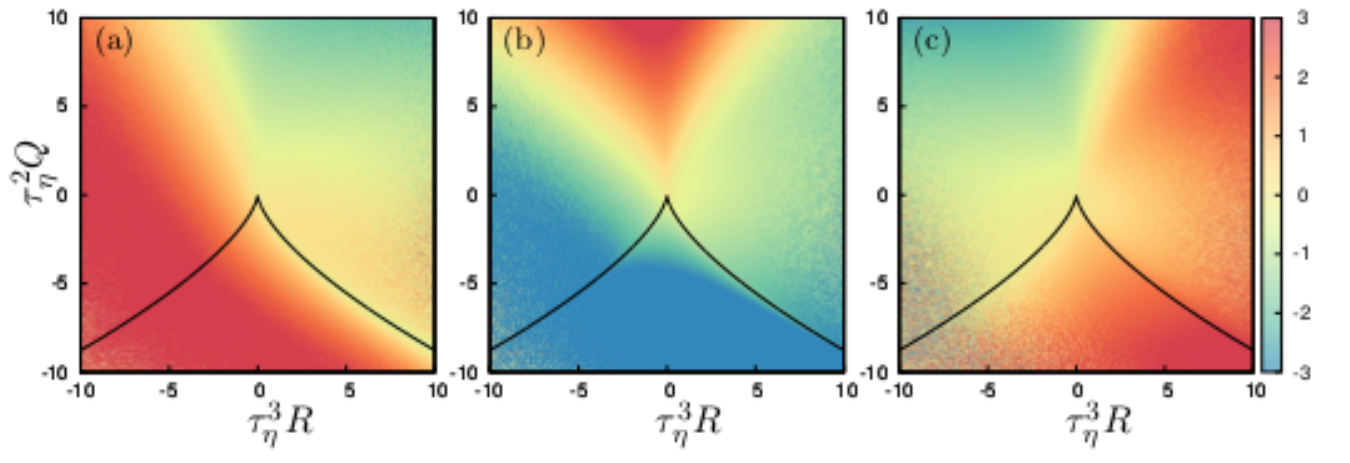}
\caption{Results for $\langle-\mathcal{\widetilde{H}}^*_{\gamma,i(i)}\big\vert R,Q\rangle$, the average of the diagonal components of $-\aH_\gamma^*$ in the strain-rate eigenframe conditioned on the principal invariants $R,Q$. (a) $i=1$, (b) $i=2$, (c) $i=3$. Black lines denote the Vieillefosse tails.
}
\label{H_gamma_eigS}
\end{figure}

In order to fully quantify the effect of $\aH_\gamma^*$, its magnitude should also be considered together with its orientation. The average of the diagonal components of $-\aH_\gamma^*$ in the strain-rate eigenframe conditioned on the invariants $R,Q$ is shown in figure \ref{H_gamma_eigS}. 
Despite the large magnitude of the rank-reduced anisotropic pressure Hessian eigenvalue, the contribution of $\aH_\gamma^*$ to the strain-rate eigenvalue dynamics is moderate on average.
Figure \ref{eigen_QR} shows that the eigenvalue of $\aH_\gamma^*$, namely $\psi$, is very large along the right Vieillefosse tail and in the quadrant $Q>0,R<0$.
Figures \ref{blow_up}(a--c) show that $\langle|\widetilde{{z}}_{i3}^2 - \widetilde{{z}}_{i1}^2|\big\vert R,Q\rangle$ is small along the right Vieillefosse tail, and these small values of $|\widetilde{z}_{i3}-\widetilde{z}_{i1}|$ compensate the large magnitude of $\psi$ in the same region.
In particular, the orientational contribution of $\aH_\gamma^*$ to the dynamics of $\lambda_1$, namely $|\widetilde{z}_{13}-\widetilde{z}_{11}|$, is very small along the right Vieillefosse tail.
This indicates how the direct amplification of $\lambda_1$ due to $\aH_\gamma^*$ does not lead to blow up, since this amplification is strong for $R<0$, but is very weak along the right Vieillefosse tail where RE blows up, as shown in figure \ref{H_gamma_eigS}(a).
As observed above, the rank-reduced anisotropic pressure Hessian tends to suppress positive values of $\lambda_2$ in the $R>0,Q<0$ quadrant, as displayed in figure \ref{H_gamma_eigS}(b). Interestingly, however, $\aH_\gamma^*$ contributes to the growth of positive $\lambda_2$ in the region $Q>0,R<0$, where $\bm{\omega}$ and $\bm{v}_2$ are also strongly aligned (see figure \ref{fig_vorticity_S_alignment_RQ} (b)). As such, $\aH_\gamma^*$ indirectly contributes to vortex stretching.
The results in figure \ref{H_gamma_eigS}(c) show that, the rank-reduced anisotropic pressure Hessian strongly hinders $\lambda_3$ along the right Vieillefosse tail, contributing to its amplification only in a small region where $R<0$ and $Q>0$. This is a key way in which $\aH_\gamma^*$ acts to prevent blow-up in the region $R>0,Q<0$.

\subsection{Rank-redued anisotropic pressure Hessian--vorticity alignment}

As shown earlier, $\aH_\gamma^*$ exhibits remarkable alignment properties with respect to the vorticity $\bm{\omega}$. In view of this, we now consider how this alignment impacts the way that $\aH_\gamma^*$ competes with the centrifugal term produced by vorticity to control the growth of the strain-rates. This can be explored by considering the strain-rates along the vorticity direction.

\begin{figure}
\includegraphics[width=1.\textwidth]{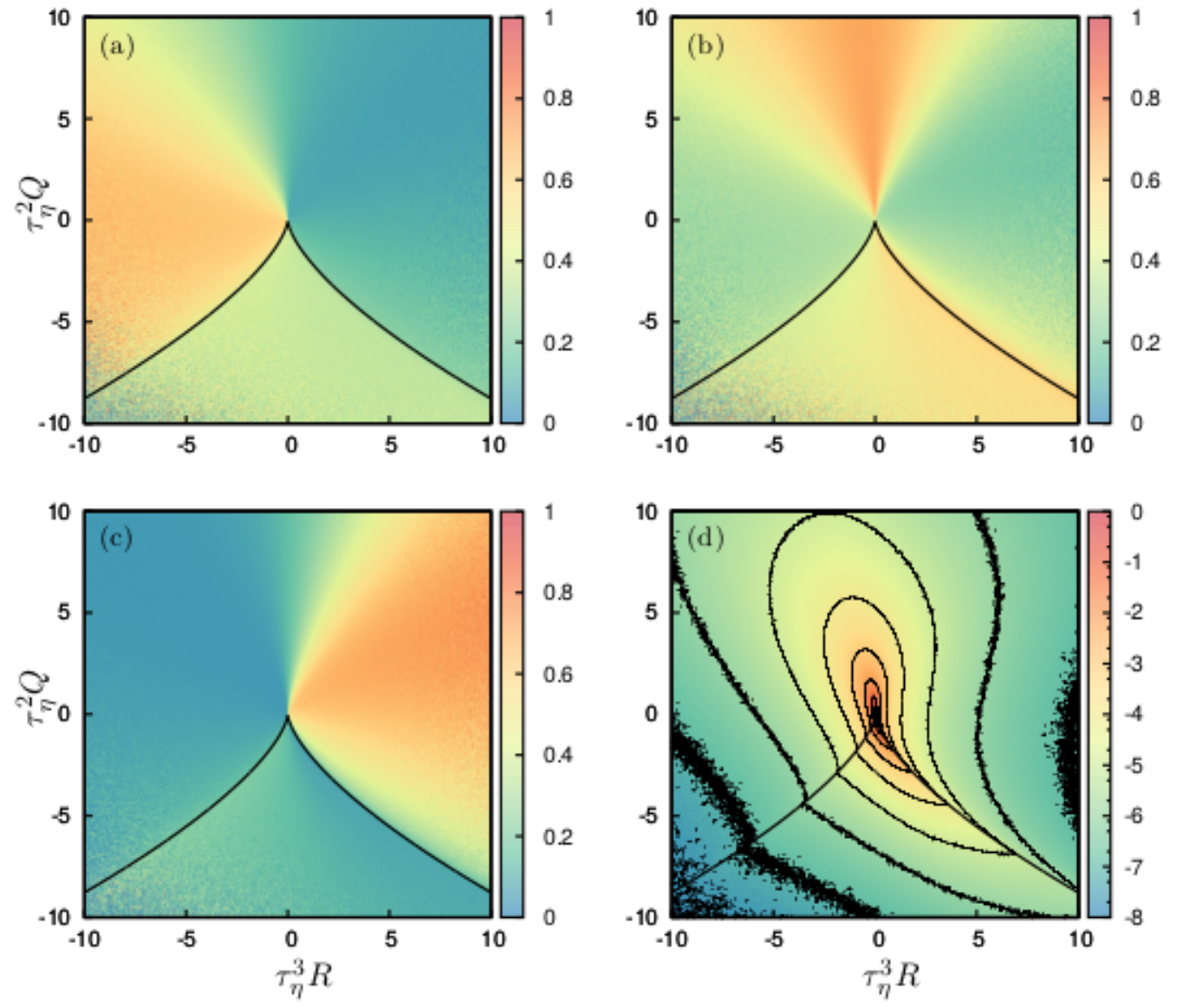}
\caption{Results for $\langle (\hat{\bm{\omega}}\bm{\cdot}\bm{v}_i)^2\big\vert R,Q\rangle$, the statistical alignment between vorticity and eigenvectors of the strain-rate tensor, conditioned on the principal invariants $R,Q$. (a) $i=1$, (b) $i=2$, (c) $i=3$. Plot (d) shows the joint probability density of the principal invariants $R$ and $Q$. Black lines denote the Vieillefosse tails.}
\label{fig_vorticity_S_alignment_RQ}
\end{figure}

The statistical alignments of the vorticity vector with the strain-rate eigenvectors, quantified by $(\bm{v}_i\bm{\cdot \hat{\omega}})^2$, conditioned on the invariants $R$ and $Q$, are shown in figure \ref{fig_vorticity_S_alignment_RQ}.
The vorticity tends to align with the most extensional strain-rate eigenvector in the region $R<0$ and also, to a lesser extent, between the Vieillefosse tails. Alignment between the vorticity and the most compressional strain-rate eigenvector takes place in the region $R>0$ only, above the right Vieillefosse tail.
The vorticity vector strongly aligns with the intermediate strain-rate eigenvector in the region $Q>0$, close to the $R=0$ axis and along the right Vieillefosse tail. 
The half-plane $Q>0$ and the vicinity of the right Vieillefosse tail correspond to the bulk of probability on the $Q,R$ plane \citep{Meneveau2011}, as shown in figure \ref{fig_vorticity_S_alignment_RQ}(d), and therefore preferential alignment between vorticity and the intermediate strain-rate eigenvector is observed. In the phase-space region in which the alignment between vorticity and the intermediate strain-rate eigenvector is strong, the contribution of $\aH_\gamma^*$ to the dynamics of $\lambda_2$ is larger. This is observed by comparing figure \ref{fig_vorticity_S_alignment_RQ}(b) and figure \ref{H_gamma_eigS}(b).

\begin{figure}
\includegraphics[width=1.\textwidth]{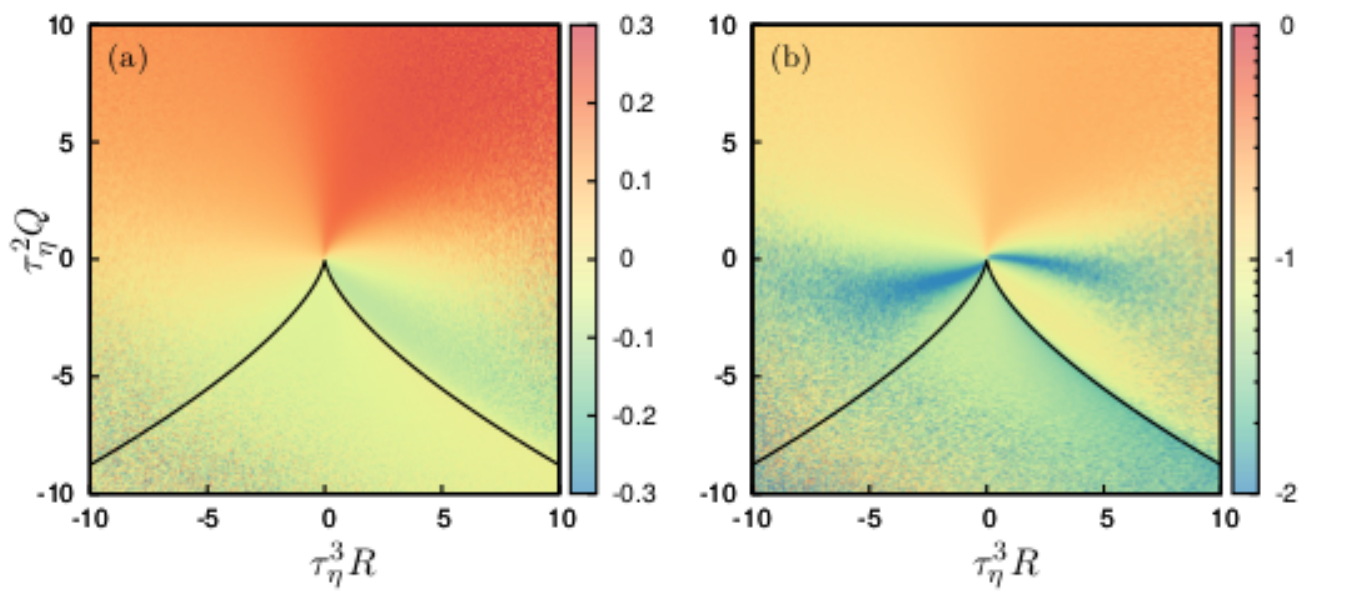}
\caption{Statistical alignment between vorticity and eigenvalues of the rank-reduced anisotropic pressure Hessian, conditioned on the principal invariants $R,Q$. Results for (a) $\langle (\hat{\bm{\omega}}\bm{\cdot}\bm{z}_3)^2-(\hat{\bm{\omega}}\bm{\cdot}\bm{z}_1)^2 \big\vert R,Q\rangle$ and (b) $\langle |(\hat{\bm{\omega}}\bm{\cdot}\bm{z}_3)^2-(\hat{\bm{\omega}}\bm{\cdot}\bm{z}_1)^2| \big\vert R,Q\rangle$ in logarithmic scale. Black lines denote the Vieillefosse tails.}
\label{fig_vorticity_H_gamma_alignment_RQ}
\end{figure}

We now turn to the combined effects of $\aH_\gamma^*$ and $\bm{\omega}$ on the strain-rate dynamics.
The evolution equation for $\bm{S}$ may be written as (ignoring the viscous term) 
\begin{equation}
D_t\bm{S}=-\left(\bm{S\cdot S}-\frac{\Tr(\bm{S\cdot S})}{3}\mathbf{I}  \right) -\frac{1}{4}\left(\bm{\omega}\bm{\omega}^\top- \frac{\omega^2}{3} \mathbf{I} \right)-\bm{\mathcal{H}}.
\end{equation}
We consider the projection of this equation along the instantaneous vorticity direction $\hat{\bm{\omega}}\equiv\bm{\omega}/\omega$, and along this direction, the contribution of the last two terms is
\begin{equation}
\hat{\bm{\omega}}\bm{\cdot} \left(-\frac{1}{4}\bm{\omega}\bm{\omega}^\top+ \frac{\omega^2}{12} \mathbf{I} -\bm{\mathcal{H}}    \right)   \bm{\cdot}\hat{\bm{\omega}}  =
-\frac{1}{6}\omega^2 -\psi\left( (\hat{\bm{\omega}}\bm{\cdot}\bm{z}_1)^2-(\hat{\bm{\omega}}\bm{\cdot}\bm{z}_3)^2 \right), 
\label{eq_H_gamma_R2}
\end{equation}
where the properties of $\aH_{\gamma}^*$ have allowed us to use $\aH_{\gamma}^*$ instead of $\aH$. Note that the term $-\omega^2/6$ comes entirely from the contribution of vorticity to the isotropic part of the pressure Hessian, since the centrifugal contribution does not act along the direction of vorticity, but only orthogonal to it.
Equation \eqref{eq_H_gamma_R2} shows that (noting $\psi\geq 0$) when the vorticity is more aligned with the extensional/compressional direction of $\aH_\gamma^*$, then $\aH_\gamma^*$ acts with/against the contribution from vorticity to oppose/aid the production of strain along the vorticity direction.
In figure \ref{fig_vorticity_H_gamma_alignment_RQ} we consider the DNS data for $\langle (\hat{\bm{\omega}}\bm{\cdot}\bm{z}_3)^2-(\hat{\bm{\omega}}\bm{\cdot}\bm{z}_1)^2 \big\vert R,Q\rangle$.
The results show that in $Q>0$ regions, the vorticity vector preferentially aligns with the most compressional eigenvector of the rank-reduced anisotropic pressure Hessian, so that $\langle (\hat{\bm{\omega}}\bm{\cdot}\bm{z}_3)^2-(\hat{\bm{\omega}}\bm{\cdot}\bm{z}_1)^2 \big\vert R,Q\rangle>0$. On the contrary, in $Q<0$ regions  $\langle (\hat{\bm{\omega}}\bm{\cdot}\bm{z}_3)^2-(\hat{\bm{\omega}}\bm{\cdot}\bm{z}_1)^2 \big\vert R,Q\rangle<0$. This striking behavior means that in vorticity dominated regions,  $\aH_\gamma^*$ acts to increase the strain-rate along the vorticity direction, and the opposite in strain dominated regions.

\section{Conclusions} 
In this paper a new symmetry for the Lagrangian dynamics of the velocity gradient invariants has been presented and it  has been interpreted as a gauge for the anisotropic pressure Hessian.
This gauge arises because the dynamics of the strain-rate eigenvalues and vorticity components in the strain-rate eigenframe are unaffected by the angular velocity of the eigenframe along the vorticity direction. Using this symmetry, we have introduced a modified pressure Hessian, $\aH_\gamma^*$, that is the sum of the standard pressure Hessian and the gauge term. We then sought for lower dimensional representations of the pressure Hessian by performing a rank-reduction on $\aH_\gamma^*$, allowed by the additional degree of freedom provided by the gauge symmetry.
Remarkably, this rank reduction is possible everywhere in the flow, and consequently everywhere in the flow a two-dimensional $\aH_\gamma^*$ may be defined that generates exactly the same eigenframe dynamics as the full three-dimensional pressure Hessian $\aH$.
We also showed that $\aH_\gamma^*$ exhibits remarkable alignment properties with respect to the strain-rate eigenframe and vorticity, that are not possessed by $\aH$.
In particular, the plane on which $\aH_\gamma^*$ acts tends to be almost orthogonal to the vorticity vector. Consistently, the intermediate eigenvector of $\aH_\gamma^*$ strongly aligns with the strain-rate intermediate eigenvector. Also, the most compressional/extensional eigenvectors of $\aH_\gamma^*$ preferentially form an angle of $\pi/4$ with the most compressional/extensional eigenvectors of the strain-rate tensor.

The rank-reduced anisotropic pressure Hessian offers promising applications. For example, the reduction in dimensionality,  provided by replacing $\aH$ with $\aH_\gamma^*$ in the eigenframe equations, is a step towards more efficient modeling, since the rank-reduced anisotropic pressure Hessian can be specified by only four numbers instead of five required for the fully three-dimensional $\aH$. The eigenvalues of $\aH_\gamma^*$ are also shown to be strongly related to the local strain-rate and vorticity in the flow, suggesting relatively simple ways to model these eigenvalues in Lagrangian models for the velocity gradient tensor. This property, together with the reduction in dimensionality and the remarkable alignment properties of $\aH_\gamma^*$, offer promising insights into ways in which the anisotropic pressure Hessian and its effects on the eigenframe dynamics can be modelled. The development of such a model will be the subject of future work.

\section*{Acknowledgements}
This work used the Extreme Science and Engineering Discovery Environment (XSEDE), supported by National Science Foundation grant ACI-1548562 \citep{xsede}.

\section{Declaration of Interests}
 The authors report no conflict of interest.

\bibliographystyle{jfm}
\bibliography{bib_gradient}

\begin{thebibliography}{33}
\expandafter\ifx\csname natexlab\endcsname\relax\def\natexlab#1{#1}\fi
\def\au#1{#1} \def\ed#1{#1} \def\yr#1{#1}\def\at#1{#1}\def\jt#1{\textit{#1}}
  \def\bt#1{#1}\def\bvol#1{\textbf{#1}} \def\vol#1{#1} \def\pg#1{#1}
  \def\publ#1{#1}\def\arxiv#1{#1}\def\org#1{#1}\def\st#1{\textit{#1}}

\bibitem[Ashurst {\em et~al.\/}(1987)Ashurst, Kerstein, Kerr \&
  Gibson]{Ashurst1987}
{\sc \au{Ashurst, W.~T.}, \au{Kerstein, A.~R.}, \au{Kerr, R.~M.} \& \au{Gibson,
  C.~H.}} \yr{1987}  \at{Alignment of vorticity and scalar gradient with strain
  rate in simulated {N}avier--{S}tokes turbulence}.  \jt{The Physics of Fluids}
   \bvol{30}~(8),  \pg{2343--2353}.

\bibitem[Ballouz \& Ouellette(2018)]{Ballouz2018}
{\sc \au{Ballouz, J.~G.} \& \au{Ouellette, N.~T.}} \yr{2018}  \at{Tensor
  geometry in the turbulent cascade}.  \jt{Journal of Fluid Mechanics}
  \bvol{835},  \pg{1048--1064}.

\bibitem[Batchelor \& Taylor(1952)]{Batchelor1952}
{\sc \au{Batchelor, G.~K.} \& \au{Taylor, G.~I.}} \yr{1952}  \at{The effect of
  homogeneous turbulence on material lines and surfaces}.  \jt{Proceedings of
  the Royal Society of London. Series A. Mathematical and Physical Sciences}
  \bvol{213}~(1114),  \pg{349--366}.

\bibitem[Betchov(1956)]{Betchov1956}
{\sc \au{Betchov, R.}} \yr{1956}  \at{An inequality concerning the production
  of vorticity in isotropic turbulence}.  \jt{Journal of Fluid Mechanics}
  \bvol{1}~(5),  \pg{497--504}.

\bibitem[Buaria {\em et~al.\/}(2019)Buaria, Pumir, Bodenschatz \&
  Yeung]{Buaria2019}
{\sc \au{Buaria, D.}, \au{Pumir, A.}, \au{Bodenschatz, E.} \& \au{Yeung,
  P.~K.}} \yr{2019}  \at{Extreme velocity gradients in turbulent flows}.
  \jt{New Journal of Physics}  \bvol{21}~(4),  \pg{043004}.

\bibitem[Cantwell(1992)]{Cantwell1992}
{\sc \au{Cantwell, B.~J.}} \yr{1992}  \at{Exact solution of a restricted
  {E}uler equation for the velocity gradient tensor}.  \jt{Physics of Fluids A:
  Fluid Dynamics}  \bvol{4}~(4),  \pg{782--793}.

\bibitem[Chertkov {\em et~al.\/}(1999)Chertkov, Pumir \&
  Shraiman]{Chertkov1991}
{\sc \au{Chertkov, M.}, \au{Pumir, A.} \& \au{Shraiman, B.~I.}} \yr{1999}
  \at{Lagrangian tetrad dynamics and the phenomenology of turbulence}.
  \jt{Physics of Fluids}  \bvol{11}~(8),  \pg{2394--2410}.

\bibitem[Chevillard \& Meneveau(2006)]{Chevillard2006}
{\sc \au{Chevillard, L.} \& \au{Meneveau, C.}} \yr{2006}  \at{Lagrangian
  dynamics and statistical geometric structure of turbulence}.  \jt{Phys. Rev.
  Lett.}  \bvol{97},  \pg{174501}.

\bibitem[Chevillard {\em et~al.\/}(2008)Chevillard, Meneveau, Biferale \&
  Toschi]{Chevillard2008}
{\sc \au{Chevillard, L.}, \au{Meneveau, C.}, \au{Biferale, L.} \& \au{Toschi,
  F.}} \yr{2008}  \at{Modeling the pressure {H}essian and viscous {L}aplacian
  in turbulence: {C}omparisons with direct numerical simulation and
  implications on velocity gradient dynamics}.  \jt{Physics of Fluids}
  \bvol{20}~(10),  \pg{101504}.

\bibitem[Dresselhaus \& Tabor(1992)]{Dresselhaus1992}
{\sc \au{Dresselhaus, E.} \& \au{Tabor, M.}} \yr{1992}  \at{The kinematics of
  stretching and alignment of material elements in general flow fields}.
  \jt{Journal of Fluid Mechanics}  \bvol{236},  \pg{415--444}.

\bibitem[Falkovich \& Gaw\c{e}dzki(2014)]{Falkovich2014}
{\sc \au{Falkovich, G.} \& \au{Gaw\c{e}dzki, K.}} \yr{2014}  \at{Turbulence on
  hyperbolic plane: {T}he fate of inverse cascade}.  \jt{Journal of Statistical
  Physics}  \bvol{156}~(1),  \pg{10--54}.

\bibitem[Girimaji \& Pope(1990)]{Girimaji1990a}
{\sc \au{Girimaji, S.~S.} \& \au{Pope, S.~B.}} \yr{1990}  \at{A diffusion model
  for velocity gradients in turbulence}.  \jt{Physics of Fluids A: Fluid
  Dynamics}  \bvol{2}~(2),  \pg{242--256}.

\bibitem[Ireland {\em et~al.\/}(2016{\natexlab{{\em a\/}}})Ireland, Bragg \&
  Collins]{Ireland2016a}
{\sc \au{Ireland, P.~J.}, \au{Bragg, A.~D.} \& \au{Collins, L.~R.}}
  \yr{2016{\natexlab{{\em a\/}}}}  \at{The effect of {R}eynolds number on
  inertial particle dynamics in isotropic turbulence. {P}art 1. {S}imulations
  without gravitational effects}.  \jt{Journal of Fluid Mechanics}  \bvol{796},
   \pg{617--658}.

\bibitem[Ireland {\em et~al.\/}(2016{\natexlab{{\em b\/}}})Ireland, Bragg \&
  Collins]{Ireland2016b}
{\sc \au{Ireland, P.~J.}, \au{Bragg, A.~D.} \& \au{Collins, L.~R.}}
  \yr{2016{\natexlab{{\em b\/}}}}  \at{The effect of {R}eynolds number on
  inertial particle dynamics in isotropic turbulence. {P}art 2. {S}imulations
  with gravitational effects}.  \jt{Journal of Fluid Mechanics}  \bvol{796},
  \pg{659--711}.

\bibitem[Ireland {\em et~al.\/}(2013)Ireland, Vaithianathan, Sukheswalla, Ray
  \& Collins]{Ireland2013}
{\sc \au{Ireland, P.~J.}, \au{Vaithianathan, T.}, \au{Sukheswalla, P.~S.},
  \au{Ray, B.} \& \au{Collins, L.~R.}} \yr{2013}  \at{Highly parallel
  particle-laden flow solver for turbulence research}.  \jt{Computers \&
  Fluids}  \bvol{76},  \pg{170--177}.

\bibitem[Johnson \& Meneveau(2016)]{Johnson2016}
{\sc \au{Johnson, P.~L.} \& \au{Meneveau, C.}} \yr{2016}  \at{A closure for
  lagrangian velocity gradient evolution in turbulence using recent-deformation
  mapping of initially {G}aussian fields}.  \jt{Journal of Fluid Mechanics}
  \bvol{804},  \pg{387--419}.

\bibitem[Johnson \& Meneveau(2017)]{Johnson2017}
{\sc \au{Johnson, P.~L.} \& \au{Meneveau, C.}} \yr{2017}  \at{Turbulence
  intermittency in a multiple-time-scale {N}avier-{S}tokes-based reduced
  model}.  \jt{Phys. Rev. Fluids}  \bvol{2},  \pg{072601}.

\bibitem[Kraichnan(1970)]{Kraichnan1970}
{\sc \au{Kraichnan, R.~H.}} \yr{1970}  \at{Diffusion by a random velocity
  field}.  \jt{The Physics of Fluids}  \bvol{13}~(1),  \pg{22--31}.

\bibitem[Lawson \& Dawson(2015)]{Lawson2015}
{\sc \au{Lawson, J.~M.} \& \au{Dawson, J.~R.}} \yr{2015}  \at{On velocity
  gradient dynamics and turbulent structure}.  \jt{Journal of Fluid Mechanics}
  \bvol{780},  \pg{60--98}.

\bibitem[Majda \& Bertozzi(2001)]{Majda2001}
{\sc \au{Majda, A.~J.} \& \au{Bertozzi, A.~L.}} \yr{2001} {\em Vorticity and
  Incompressible Flow\/}.  \publ{Cambridge University Press}.

\bibitem[Meneveau(2011)]{Meneveau2011}
{\sc \au{Meneveau, C.}} \yr{2011}  \at{Lagrangian dynamics and models of the
  velocity gradient tensor in turbulent flows}.  \jt{Annual Review of Fluid
  Mechanics}  \bvol{43}~(1),  \pg{219--245}.

\bibitem[Naso \& Pumir(2005)]{Naso2005}
{\sc \au{Naso, A.} \& \au{Pumir, A.}} \yr{2005}  \at{Scale dependence of the
  coarse-grained velocity derivative tensor structure in turbulence}.
  \jt{Phys. Rev. E}  \bvol{72},  \pg{056318}.

\bibitem[Nomura \& Post(1998)]{Nomura1998}
{\sc \au{Nomura, K.~K.} \& \au{Post, G.~K.}} \yr{1998}  \at{The structure and
  dynamics of vorticity and rate of strain in incompressible homogeneous
  turbulence}.  \jt{Journal of Fluid Mechanics}  \bvol{377},  \pg{65--97}.

\bibitem[Ohkitani(1993)]{Ohkitani1993}
{\sc \au{Ohkitani, K.}} \yr{1993}  \at{Eigenvalue problems in three-dimensional
  {E}uler flows}.  \jt{Physics of Fluids A: Fluid Dynamics}  \bvol{5}~(10),
  \pg{2570--2572}.

\bibitem[Pope(2000)]{pope}
{\sc \au{Pope, S.~B.}} \yr{2000} {\em Turbulent Flows\/}.  \publ{New York:
  Cambridge University Press}.

\bibitem[Towns {\em et~al.\/}(2014)Towns, Cockerill, Dahan, Foster, Gaither,
  Grimshaw, Hazlewood, Lathrop, Lifka, Peterson, Roskies, Scott \&
  Wilkins-Diehr]{xsede}
{\sc \au{Towns, J.}, \au{Cockerill, T.}, \au{Dahan, M.}, \au{Foster, I.},
  \au{Gaither, K.}, \au{Grimshaw, A.}, \au{Hazlewood, V.}, \au{Lathrop, S.},
  \au{Lifka, D.}, \au{Peterson, G.~D.}, \au{Roskies, R.}, \au{Scott, J.~R.} \&
  \au{Wilkins-Diehr, N.}} \yr{2014}  \at{{XSEDE}: {A}ccelerating {S}cientific
  {D}iscovery}.  \jt{Computing in Science \& Engineering}  \bvol{16}~(5),
  \pg{62--74}.

\bibitem[Tsinober(2001)]{Tsinober2001}
{\sc \au{Tsinober, A.}} \yr{2001} {\em An informal introduction to
  turbulence\/}.  \publ{Kluwer Academic Publishers}.

\bibitem[Vieillefosse(1982)]{Vieillefosse1982}
{\sc \au{Vieillefosse, P.}} \yr{1982}  \at{Local interaction between vorticity
  and shear in a perfect incompressible fluid}.  \jt{J. Phys. France}
  \bvol{43}~(6),  \pg{837--842}.

\bibitem[Vieillefosse(1984)]{Vieillefosse1984}
{\sc \au{Vieillefosse, P.}} \yr{1984}  \at{Internal motion of a small element
  of fluid in an inviscid flow}.  \jt{Physica A: Statistical Mechanics and its
  Applications}  \bvol{125}~(1),  \pg{150--162}.

\bibitem[Vlaykov \& Wilczek(2019)]{Vlaykov2019}
{\sc \au{Vlaykov, D.~G.} \& \au{Wilczek, M.}} \yr{2019}  \at{On the small-scale
  structure of turbulence and its impact on the pressure field}.  \jt{Journal
  of Fluid Mechanics}  \bvol{861},  \pg{422--446}.

\bibitem[Wilczek \& Friedrich(2009)]{Wilczek2009}
{\sc \au{Wilczek, M.} \& \au{Friedrich, R.}} \yr{2009}  \at{Dynamical origins
  for non-gaussian vorticity distributions in turbulent flows}.  \jt{Phys. Rev.
  E}  \bvol{80},  \pg{016316}.

\bibitem[Wilczek \& Meneveau(2014)]{Wilczek2014}
{\sc \au{Wilczek, M.} \& \au{Meneveau, C.}} \yr{2014}  \at{Pressure hessian and
  viscous contributions to velocity gradient statistics based on {G}aussian
  random fields}.  \jt{Journal of Fluid Mechanics}  \bvol{756},  \pg{191--225}.

\bibitem[Yeung {\em et~al.\/}(2012)Yeung, Donzis \& Sreenivasan]{Yeung2012}
{\sc \au{Yeung, P.~K.}, \au{Donzis, D.~A.} \& \au{Sreenivasan, K.~R.}}
  \yr{2012}  \at{Dissipation, enstrophy and pressure statistics in turbulence
  simulations at high {R}eynolds numbers}.  \jt{Journal of Fluid Mechanics}
  \bvol{700},  \pg{5--15}.

\end{thebibliography}

\end{document}